\documentclass[aps,reprint,10pt,superscriptaddress,showpacs]{revtex4-1}

\usepackage{amsmath,amssymb,amsthm,amscd,latexsym,epsfig,bm,subfigure}
\usepackage[usenames,dvipsnames]{color}
\usepackage[colorlinks,bookmarks=false,citecolor=blue,linkcolor=red,urlcolor=blue,breaklinks=true]{hyperref}
\usepackage{verbatim}
\usepackage{bbm}
\usepackage{array}
\usepackage[normalem]{ulem}

\def\beq{\begin{equation}}
\def\eeq{\end{equation}}
\def\be{\begin{equation}}
\def\ee{\end{equation}}

\newcommand{\zz}{\mathbb{Z}_2}
\newcommand{\z}{\mathbb{Z}}

\def\r{{\bf r}}

\def\br{\boldsymbol{r}}

\def\f2{{\mathbb F}_2}

\def\tC{\tilde{\cal C}}

\theoremstyle{plain}
\theoremstyle{plain}

\providecommand{\theoremname}{Theorem}
\providecommand{\theoremtextname}{Theorem}

\theoremstyle{plain}
\providecommand{\propositionname}{Proposition}

\newcommand{\PreserveBackslash}[1]{\let\temp=\\#1\let\\=\temp}
\newcolumntype{C}[1]{>{\PreserveBackslash\centering}p{#1}}
\newcolumntype{R}[1]{>{\PreserveBackslash\raggedleft}p{#1}}
\newcolumntype{L}[1]{>{\PreserveBackslash\raggedright}p{#1}}

\begin{document}

\title{Topological states from topological crystals}

\author{Zhida Song}
\affiliation{Beijing National Research Center for Condensed Matter Physics,
and Institute of Physics, Chinese Academy of Sciences, Beijing 100190, China}
\affiliation{University of Chinese Academy of Sciences, Beijing 100049, China}
\author{Sheng-Jie Huang}
\affiliation{Department of Physics, University of Colorado, Boulder, Colorado 80309, USA}
\affiliation{Center for Theory of Quantum Matter, University of Colorado, Boulder, Colorado 80309, USA}
\author{Yang Qi}
\affiliation{Center for Field Theory and Particle Physics, Department of Physics, Fudan University, Shanghai 200433, China}
\affiliation{State Key Laboratory of Surface Physics, Fudan University, Shanghai 200433, China}
\affiliation{Collaborative Innovation Center of Advanced Microstructures, Nanjing 210093, China}
\author{Chen Fang}
\email{cfang@iphy.ac.cn}
\affiliation{Beijing National Research Center for Condensed Matter Physics,
and Institute of Physics, Chinese Academy of Sciences, Beijing 100190, China}
\affiliation{CAS Center for Excellence in Topological Quantum Computation, Beijing, China}
\author{Michael Hermele}
\email{michael.hermele@colorado.edu}
\affiliation{Department of Physics, University of Colorado, Boulder, Colorado 80309, USA}
\affiliation{Center for Theory of Quantum Matter, University of Colorado, Boulder, Colorado 80309, USA}
\date{\today}

\begin{abstract}
We present a scheme to explicitly construct and classify general topological states jointly protected by an onsite symmetry group and a spatial symmetry group.
We show that all these symmetry protected topological states can be adiabatically deformed (allowing for stacking of trivial states) into a special class of states we call \textit{topological crystals}.
A topological crystal in, for example, three dimensions is a real-space assembly of finite-sized pieces of topological states in one and two dimensions protected by the local symmetry group alone, arranged in a configuration invariant under the spatial group and glued together such there is no open edge or end.
As a demonstration of principle, we explicitly enumerate all inequivalent topological crystals for non-interacting time-reversal symmetric electronic insulators with significant spin-orbit coupling and any one of the 230 space groups in three dimensions.
Because every topological crystalline insulator can be deformed into a topological crystal, the enumeration of the latter gives topological crystalline insulators a full classification and for each class an explicit  real-space construction.  We also extend these results to give a unified classification including both strong topological insulators and topological crystalline insulators.
\end{abstract}

\maketitle

\section*{Introduction} 

Symmetry protected topological (SPT) phases are gapped many-body ground states that can only be adiabatically deformed into product states of local orbitals by breaking a given symmetry group or by closing the energy gap \cite{Schnyder2008, Kitaev2009, ryu2010, gu2009, pollmann2010, fidkowski2011, turner2011, chen2011_1dspt, chen2011_1dcomplete, cirac2011, chen2013cohomology, levin2012}.
Typical examples are topological insulators, topological superconductors and the Haldane spin chain \cite{Hasan2010,Qi2011,turner2013topological,Haldane1983}.
The best-understood SPT phases are those of non-interacting fermions. 
Not long after the discovery of topological band insulators, free-fermion topological phases were completely classified for systems with internal (\textit{i.e.} non-spatial) symmetries \cite{Schnyder2008,Kitaev2009,ryu2010,Chiu2016}. 
Of course, crystalline symmetries play a central role in solid state physics, so attention naturally began to turn to topological crystalline insulators (TCIs), which are electronic insulators whose topologically non-trivial nature is protected, in part, by point group or space group symmetry\cite{Fu2011,Hsieh2012,ando2015,Chiu2016}. 
Sparked by the prediction and observation of TCIs in SnTe\cite{Hsieh2012,Dziawa2012,Tanaka2012,Xu2012},
%and KHgSb\cite{Wang2016,Ma2017},
remarkable theoretical\cite{Chiu2013,Liu2014,Shiozaki2014,Fang2015,Shiozaki2015,Shiozaki2016,Wang2016,Bradlyn2017,Po2017,Fang2017a,kruthoff2017,Wieder2018} and experimental progress\cite{Okada2013,Sessi2016,Ma2017} has followed over the last few years.

Despite these developments, somewhat surprisingly, a unified picture of the classification of non-interacting electron TCIs has yet to emerge.
The primary tool for the classification of free-fermion topological phases with spatial symmetries has been $K$-theory\cite{Kitaev2009} and equivariant $K$-theory\cite{Freed2013,Shiozaki2018}. 
A number of concrete classification results have been obtained\cite{Chiu2013,Shiozaki2014,Shiozaki2015,Shiozaki2016,kruthoff2017}, but, reflecting the complexity of $K$-theory, there is a paucity of concrete results for three-dimensional ($d=3$) insulators with general space group symmetry and time-reversal symmetry.
Moreover, it is not understood how or whether electron interactions can be included within $K$-theory. 
Therefore, there is a need to develop alternate means to classify TCIs and other crystalline SPT (cSPT) phases jointly protected by internal and spatial symmetries. 
Ideally, in order to provide a useful complement to $K$-theory, such methods will be real-space-based, physically transparent, and allow for interactions to be included.

In this paper, we propose a general method for classifying and constructing cSPT phases, which is then applied to the case of electronic TCIs in all 230 space groups, with time reversal symmetry and significant spin-orbit coupling. We also extend these results to include strong topological insulators.
The key idea is to first argue that any cSPT phase is adiabatically connected to a real-space crystalline pattern of lower-dimensional topological states, which we refer to as a \textit{topological crystal} \cite{song17topological, huang17building}.
One then develops a classification of phases of matter in terms of topological crystals.
Our approach is based on recent developments in the seemingly harder problem of classifying \textit{interacting} cSPT phases\cite{song17topological,jiang17anyon,thorngren18gauging,huang17building,Han2018}.
For bosonic cSPT phases with only space group symmetry, the resulting classification\cite{huang17building} agrees with that obtained in complementary approaches based on tensor network states and gauging crystalline symmetry \cite{jiang17anyon,thorngren18gauging}.

Our approach is related to, but goes beyond, layer constructions of TCIs\cite{Isobe2015,Fulga2016,huang17building,song17topological,Fang2017a,Song2017a}. 
Indeed, any construction of a TCI in terms of decoupled layers, including the archetype of weak topological insulators as stacks of $d=2$ topological insulators, is a topological crystal. 
However, by comparison to the recent systematic study of layer constructions in~\cite{Song2017a}, we show that in certain nonsymmorphic space groups there are TCIs that do not have a layer construction, but can still be realized as topological crystals.

The results we obtain are related to recent work of Khalaf \emph{et. al.}, who considered TCIs with anomalous surface states (dubbed sTCIs), and proposed a classification for sTCIs with point group and space group symmetry via the surface states of doubled strong topological insulators \cite{Khalaf17symmetry}.  The TCI classifications produced by our approach, which focuses on the bulk and does not assume anomalous surface states or a description in terms of Dirac fermions, agree with the sTCI classifications of \cite{Khalaf17symmetry}. This agreement shows that all the TCIs we classify have anomalous surface states for some surface termination.

\section*{Topological crystals}

We begin by considering a $d=3$ system with symmetry $G = G_{{\rm int}} \times G_c$, where $G_{{\rm int}}$ is some internal symmetry, and $G_c$ is either a crystalline site symmetry (\textit{i.e.}, point group) or space group
\footnote{The assumption that $G$ is a direct product of $G_{{\rm int}}$ and $G_c$ is not necessary, and is only made for simplicity of discussion, and because it holds for the electronic TCIs to be later discussed.}.
We assume the system is in an SPT phase (which could be the trivial phase); that is, below an energy gap, the ground state $| \psi \rangle$ is unique and symmetry-preserving, and, moreover, $|\psi\rangle$ is adiabatically connected to a trivial product state if the symmetry $G$ is broken explicitly.
Moreover, we restrict to those SPT phases that only remain non-trivial in the presence of crystalline symmetry; that is, $|\psi\rangle$ is adiabatically connected to a product state if $G_c$ is explicitly broken, even if $G_{{\rm int}}$ is preserved.
To avoid complications associated with gapless boundary states, we consider periodic boundary conditions.  

\begin{figure}
\begin{centering}
\includegraphics[width=0.9\linewidth]{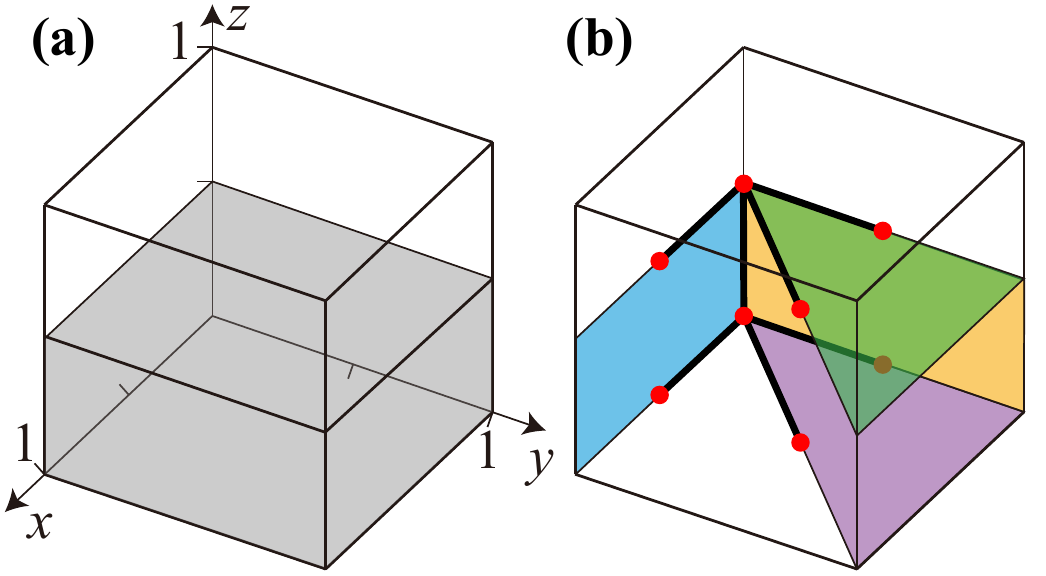}
\par\end{centering}
\caption{Cell complex structure for space group $P\bar{1}$ (\#2).  (a) The asymmetric unit $0 < x,y < 1$ and $0 < z < \frac{1}{2}$.  (b)  The symmetry-inequivalent 2-cells (colored faces), 1-cells (bold lines) and 0-cells (red dots).  The other cells can be obtained from these by acting with symmetry operations.}\label{fig:SG2}
\end{figure}

To proceed, we identify an \emph{asymmetric unit} (AU), which is the interior of a region of space that is as large as possible, subject to the condition that no two points in the region are related by a crystalline symmetry.
The AU is then copied throughout space using the crystalline symmetry, and we denote the resulting union of non-overlapping AUs as ${\cal A}$.
This construction gives three-dimensional space a cell complex structure (Appendix~\ref{app:cell_complex}), where the 3-cells are the individual (non-overlapping) copies of the AU in ${\cal A}$. 
The 2-cells lie on faces where two 3-cells meet, with the property that no two distinct points in the same 2-cell are related by symmetry. 
Similarly, 1-cells are edges where two or more faces meet, and 0-cells are points where edges meet . 
The 3-cells are in one-to-one correspondence with elements of $G_c$: arbitrarily choosing one 3-cell to correspond to the identity element, each other 3-cell is the unique image of this one upon acting with $g \in G_c$. 
The 2-skeleton $X^2$ is the complement of ${\cal A}$.   An example of this cell structure for the space group $P\bar{1}$ is shown in Fig.~\ref{fig:SG2}.

Ref.~\cite{huang17building} argued that $| \psi\rangle$ is adiabatically connected to a product of a trivial state on ${\cal A}$, with a possibly non-trivial state on $X^2$ (see Sec. VI of Ref.~\cite{huang17building}).
More precisely, one considers a thickened version of $X^2$, with characteristic thickness $w$, and its complement. 
In order for the argument to go through, it is important that $w \gg \xi$, where $\xi$ is any characteristic correlation or entanglement length of the short-range entangled state $|\psi \rangle$. 
If $G_c$ is a point group symmetry, this requires no assumptions, because $w$ can be taken sufficiently large. 
For space group symmetry, $w$ is limited by the unit cell size, and one must make the assumption that, by adding a fine mesh of trivial degrees of freedom, it is possible to make the correlation length $\xi$ as small as desired. 
This assumption not only allows the reduction to a topological crystal state, but also implies that the correlation length of the topological crystal state itself is much smaller than the unit cell size; this is important, because it allows us to associate a well-defined lower-dimensional state with each cell of $X^2$. 
While we believe this assumption is likely to hold, it is not proven, and strictly speaking it should be treated as a conjecture. 
If this conjecture is false, then our approach is simply restricted to those  cSPT phases whose correlation length is not bounded below upon adding trivial degrees of freedom.
We note that a preliminary version of the idea of reduction to $X^2$ was discussed in~\cite{huang17building}; there, unlike in the present work, the idea was not developed into a tool to obtain classifications.

The result of this reduction procedure is a topological crystal state.
The state on $X^2$ can be understood by associating a $d_b$-dimensional topological phase with each $d_b$-cell of $X^2$, where $d_b = 0,1,2$.
These lower-dimensional states are referred to as the ``building blocks'' of the topological crystal, and $d_b$ is the block dimension.
The building blocks must be glued together so as to eliminate any gapless modes in the bulk while preserving symmetry; for instance, $d_b = 2$ blocks will generally have gapless edge modes, which must gap out at the 1-cells and 0-cells where the building blocks meet.
Whereas crystals are formed by periodically arranged atoms, \emph{i.e.} zero-dimensional objects, topological crystals are ``stacked'' from building blocks which are themselves topological states in lower dimensions.

\section*{TCI Classification}

First we note that a number of topological invariants are already known that distinguish different phases and should be a part of any classification of TCIs.  In particular, these invariants include
 the weak $\mathbb{Z}_2$ invariants \cite{Fu2007,Moore2007}, mirror Chern numbers \cite{Teo2008} ($\mathbb{Z}$ invariant), and $\mathbb{Z}_2$ invariants associated with rotation \cite{Song2017}, glide-reflection \cite{Shiozaki2016,Wang2016}, inversion \cite{Turner2010,Hughes2011,Fang2017a}, roto-reflection \cite{Song2017a,Khalaf17symmetry}  and screw-rotation \cite{Song2017a,Khalaf17symmetry} symmetries. 
All possible combinations of these invariants that can be realized in TCIs with a layer construction were enumerated in \cite{Song2017a}.  
While one could not prove \emph{a priori} that these seven quantities exhaust all independent topological invariants, in this work we show they play a special role as a complete list of invariants for TCIs.
That is, we find that any two inequivalent TCIs differ by at least one of these invariants.

To further apply the tool of topological crystals to the case of electronic TCIs, we consider a system of non-interacting electrons with significant spin-orbit coupling, with internal symmetries of charge conservation and time reversal; that is, $G_{{\rm int}} = {\rm U}(1) \rtimes \zz^T$.
In addition, we have to specify the action of symmetry on fermionic degrees of freedom; for instance, we have Kramers time reversal $T$ with $T^2 = (-1)^F$, where $(-1)^F$ is the fermion parity operator.
More generally, some equations in the group $G_c$ are also modified by factors of fermion parity, in a manner determined from the $d=3$ Dirac Hamiltonian describing relativistic electrons (Appendix~\ref{app:relativistic}).
Formally, this amounts to specifying an element $\omega_f \in H^2(G, \mathbb{Z}_2)$; we emphasize that $\omega_f$ is uniquely determined by $G$ in the physical setting we are considering.

The next step is to understand what kind of topological crystal states can be placed on $X^2$.
First, we consider topological crystals built out of $d=2$ topological states.
There are two kinds of 2-cells, those that coincide with a mirror plane, and those that do not.
2-cells coinciding with a mirror plane can host a $d=2$ mirror Chern insulator (MCI) state, which is characterized by a $\z$ invariant\cite{Teo2008}.
The MCI state can be understood by diagonalizing the mirror operation $\sigma:\;z\rightarrow-z$, where $z$ is the coordinate along the normal direction of the mirror plane.
Because $\sigma^2 = (-1)^F$, one-electron wave functions can be divided into two sectors with mirror eigenvalue $\pm i$.
Because $\sigma T = T \sigma$, time reversal exchanges these two sectors, which therefore have equal and opposite Chern numbers, leading to a $\z$ invariant.
Each sector can be understood as a $d=2$ fermion system in class A, which has a $\z$ classification.
We see that the relevant symmetry class is thus effectively modified from AII to A on a mirror plane; this modification of symmetry class is familiar from classifications of reflection-symmetric free-fermion topological phases in momentum space \cite{Chiu2013}.
For 2-cells not coinciding with a mirror plane, the symmetry class remains AII, and such cells can host a $d=2$ topological insulator (2dTI), which is characterized by a $\zz$ invariant.

We also have to consider the possibility of topological crystals built from $d=1$ and $d=0$ states.
1-cells only host trivial states:
%so there are no non-trivial topological crystals with $d=1$ building blocks.
the effective symmetry class on a 1-cell can be either AII or A (see the discussion of MCIs above), and in either case the classification in $d=1$ is trivial.
On the other hand, there are non-trivial topological crystals built from $d_b = 0$ building blocks, which are atomic insulators formed from patterns of localized filled orbitals. % transforming in different irreducible representations of the site symmetry.
%Even though there are distinct atomic insulators constituting different quantum phases of matter, and even though these distinctions may be a source of interesting physics, all these states, being product states of localized orbitals, are in a sense topologically trivial.
These states are in a sense topologically trivial, and we ignore distinctions among atomic insulators in our classification.
Formally, this is accomplished by taking a certain quotient (Appendix~\ref{app:formal}.)

We thus see that there are two kinds of TCIs, both built from $d_b = 2$ blocks.
We refer to TCIs built from MCI blocks as mirror TCIs (MTCIs), while TCIs built from 2dTI blocks are dubbed $\zz$ TCIs.
Of course, a general TCI can have mixed MTCI and $\zz$ TCI character, and the classification is a product of MTCI and $\zz$ TCI classifications.
To proceed, we consider the requirement that the building blocks must be glued together to eliminate any gapless modes in the bulk.  
As shown in Appendix~\ref{app:MCIgluing}, this requirement implies that MTCIs can always be decomposed into decoupled planar MCI layers.  

\begin{figure}
\begin{centering}
\includegraphics[width=0.9\linewidth]{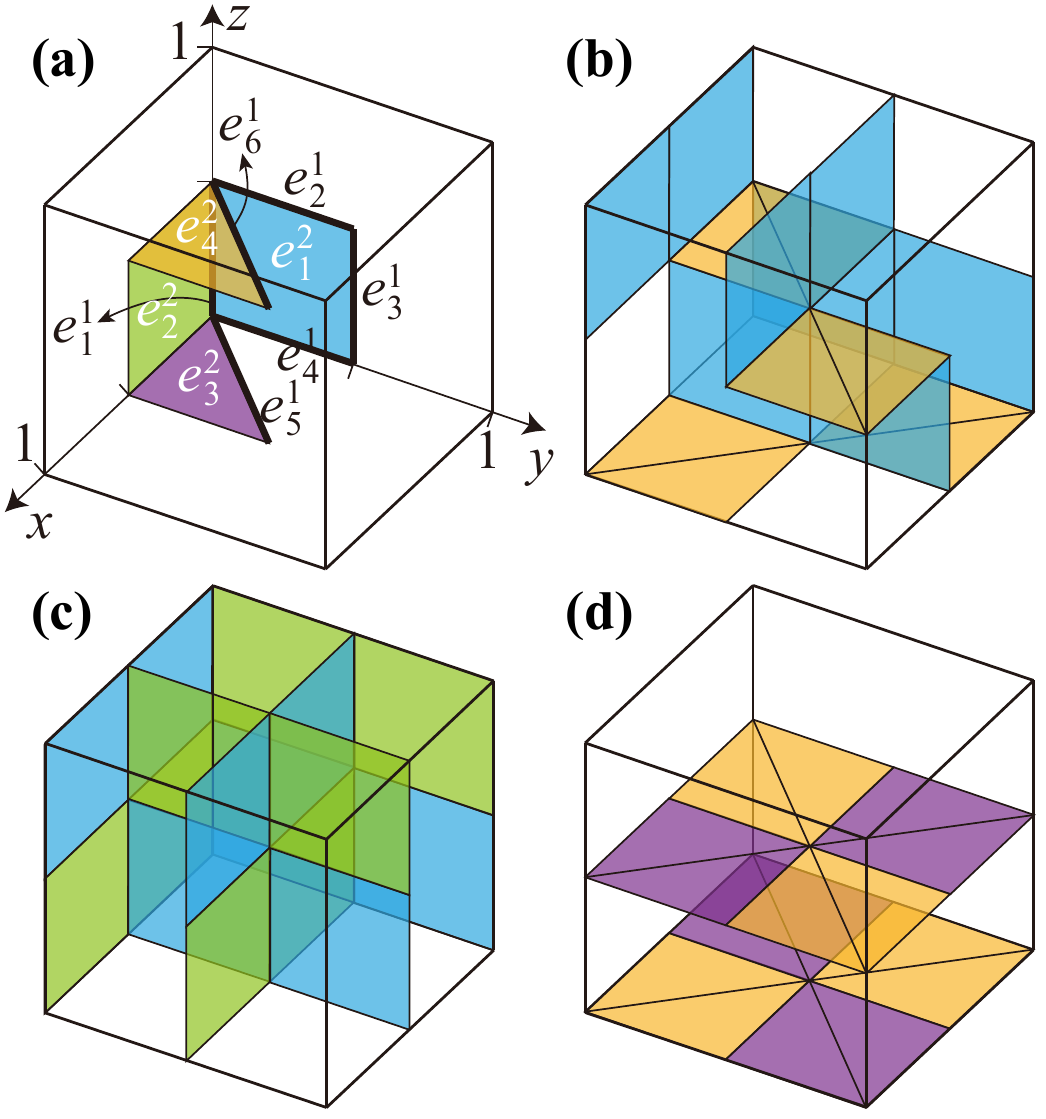}
\par\end{centering}
\caption{Topological crystals in space group $P4_22_12$ (\#94). (a) The symmetry-inequivalent 2-cells ($e^2_{i=1,2,3,4}$) and 1-cells ($e^1_{i=1,2,3,4,5,6}$) are represented by colored faces and bold lines, respectively. Here the lattice constants are set to 1, the unit cell is given by $0\le x,y,z<1$, and the AU is given by $0<x,y,z<\frac12$. (b)-(d) The three independent $\zz$ topological crystal generators, where only 2-cells decorated with 2dTIs are shown. (c) and (d) are layer constructions, whereas (b) is not. }\label{fig:PC94}
\end{figure}

$\zz$ TCIs are not quite as simple.
If we consider placing a 2dTI on some subset of the 2-cells of $X^2$, it can be shown these building blocks can be glued together into a topological crystal if and only if every 1-cell is the edge of an even number of 2dTI blocks (Appendix~\ref{app:Z2gluing}).
While sometimes such states can be decomposed into decoupled 2dTI layers, this is not always true.
For example, in space group $P4_22_12$ (\#94), for which the possible topological crystals are described below, we find a topological crystal that is beyond the scope of layer construction, as shown in Fig. \ref{fig:PC94}(b).
In this state, the 2-cells decorated with 2dTIs form a complicated yet connected structure.
Intuitively, one may lower the two yellow facets at $z=\frac12$ down to $z=0$ such that the 2dTIs form decoupled layers; however, such a process breaks the screw symmetry $\{4_{001}|\frac12\frac12\frac12\}$.
More rigorously, the non-layer-constructability can be proved by observing that the topological invariants of this state, specifically its non-trivial weak $\zz$ invariants, cannot be obtained in any TCI constructed from decoupled two-dimensional layers \cite{Song2017a}.

Having described TCIs in terms of topological crystals, we next use these states to classify TCIs.
First, we discuss equivalence relations among topological crystal states, and argue that two distinct topological crystals on $X^2$ are in different phases.
Following~\cite{song17topological, huang17building}, we need to consider an additional equivalence relation, beyond those for the $d=2$ phases of matter on the 2-cells.
Within an AU and all its copies under symmetry, we create a small bubble of 2dTI, and expand the bubble until it joins with the AU boundary; this process can be implemented with a finite-depth symmetry-preserving quantum circuit, so any two states related in this way belong to the same phase.
The reason we consider a bubble of 2dTI and not something else is that this is the only non-trivial $d=2$ state that can exist within the AU, where the only symmetries are charge conservation and time reversal.
The source of this equivalence relation is the arbitrary width $w$ of the thickened $X^2$ space in the dimensional reduction procedure; making $w$ larger corresponds to bringing in additional degrees of freedom from the ``bulk.''
However, in the present case, this equivalence operation has a trivial effect, because every 2-cell is joined with two layers of 2dTI, one on each side of the 2-cell.

Therefore, any two distinct topological crystal states are in different quantum phases of matter.
So to obtain a classification of TCIs, we need to enumerate possible topological crystals.
First we observe that topological crystals form an Abelian group ${\cal C}$ under stacking, \textit{i.e.} upon superposing two different states in the same space.  
Because MCIs (2dTIs) are characterized by $\z$ ($\zz$) invariants, ${\cal C}$ is a product of $\z$ and $\zz$ factors, with the $\z$ factors generated by MTCIs, and the $\zz$ factors generated by $\zz$ TCIs.
Because the MTCIs can be decomposed into decoupled planar layers, there is one $\z$ factor for each symmetry inequivalent set of mirror planes.
For any particular crystalline symmetry of interest, the classification ${\cal C}$ is easily worked out by considering possible colorings of the faces of the AU with MCI and 2dTI states.

To provide a concrete illustration,  we here explicitly work out the topological crystals for space group $P4_22_12$.  
$P4_22_12$ has a tetragonal lattice and is generated from three translations $\{1|100\}$, $\{1|010\}$, $\{1|001\}$, a four-fold screw $\{4_{001}|\frac12\frac12\frac12\}$, and a two-fold rotation $\{2_{110}|000\}$. (Here the lattice constants are set to 1.)
The AU can be chosen as the region $0<x,y,z<\frac12$.  The 2-cells and 1-cells are given respectively by $e^2_{i=1,2,3,4}$ and $e^1_{i=1,2,3,4,5,6}$, as shown in Fig. \ref{fig:PC94}(a), and their images under symmetry.  There are no mirror planes, so each inequivalent 2-cell can be decorated with a 2dTI state, and possible configurations are described by four $\zz$ numbers, $n_{i=1,2,3,4}$ indicating whether the corresponding $e^2_i$'s are decorated (=1) or not (=0).  The gluing condition can be expressed in a matrix form 
\beq
\sum_j A_{ij} n_j = 0 \mod 2, \label{eq:glue}
\eeq
where  $A_{i j}$ is defined to be the number of 2-cells (modulo 2) that are symmetry-equivalent to $e^2_j$ for which $e^1_i$ is an edge.  
For the setting in Fig. \ref{fig:PC94}, one can immediately read out
\beq
A = \begin{pmatrix}
0 & 0 & 0 & 0\\
1 & 1 & 1 & 1\\
0 & 0 & 0 & 0\\
1 & 1 & 1 & 1\\
0 & 0 & 0 & 0\\
0 & 0 & 0 & 0
\end{pmatrix}.
\eeq
Solving Eq. (\ref{eq:glue}), we get three independent states that generate all possible topological crystals under stacking: (i) $n_1=n_4=1$, $n_2=n_3=0$ (Fig. \ref{fig:PC94}(b)), (ii) $n_1=n_2=1$, $n_3=n_4=0$ (Fig. \ref{fig:PC94}(c)), (iii) $n_3=n_4=1$, $n_1=n_2=0$ (Fig. \ref{fig:PC94}(d)).
While states (ii) and (iii) are obviously layer constructions, state (i) is not layer-constructible, as discussed above.

\begin{figure*}
\begin{centering}
\includegraphics[width=1\linewidth]{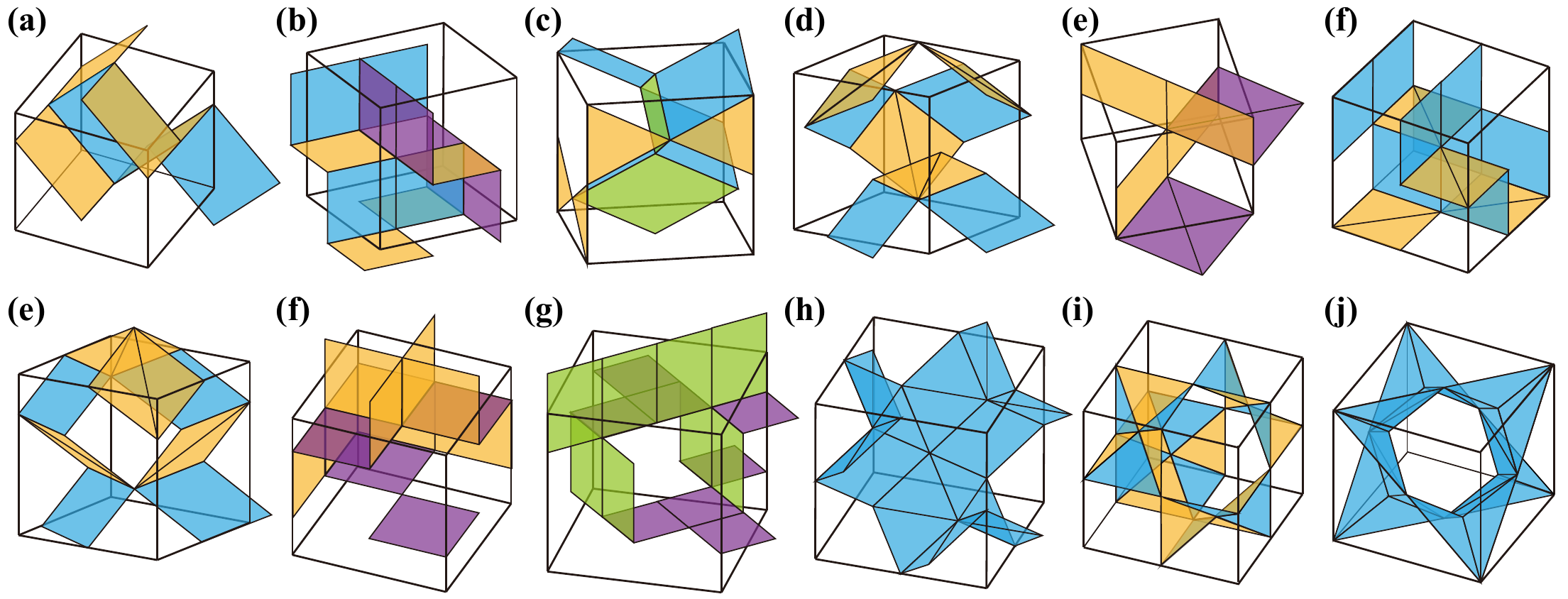}
\par\end{centering}
\caption{(a)-(j) The topological crystals beyond layer construction in space group $Pnn2$ (\#34), $Pnnn$ (\#48), $P4_2$ (\#77), $P4_2/n$ (\#86), $P4_2 22$ (\#93), $P4_2 2_1 2$ (\#94), $P4_2 cm$ (\#102), $P\bar{4}n2$ (\#118), $P4_2/nnm$ (\#134), $Pn\bar{3}$ (\#201), $P4_2 3 2$ (\#208), and $Pn\bar{3}m$ (\#224), respectively, where inequivalent 2-cells are represented by different colors. The topological invariants of these topological crystals and the coordinates of the plotted 2-cells can be found in Appendix~\ref{app:beyond}. }\label{fig:nonLC}
\end{figure*}

Now we turn to the topological invariants characterizing topological crystals.  
First, all MTCIs are characterized by real-space Chern numbers associated with certain mirror planes, and the mirror Chern numbers in momentum space for each of them are listed in \cite{Song2017a}.
Therefore, we focus on $\zz$ TCIs. 
Given a $\zz$ TCI and its corresponding topological crystal, for each symmetry operation $g \in G_c$, we assign a $\zz$ number $\delta(g)$.
We arbitrarily choose one AU and let $\r$ be a point inside, then we set $\delta(g)=1$ ($\delta(g) = 0$) if a path connecting $\r$ to $g \r$ crosses through an odd (even) number of 2dTI 2-cells.  It is shown in Appendix~\ref{app:H1} that (1) $\delta(g)$ is well-defined, independent of the arbitrary choices of AU, $\r$, and the path connecting $\r$ to $g \r$, and (2) $\delta(g_1 g_2) = \delta(g_1) + \delta(g_2)$.  The latter property implies that $\delta$ is a homomorphism from $G_c$ to $\zz$ (or, equivalently, an element of $H^1(G_c, \zz)$), which means that it is enough to specify $\delta(g)$ for the generators of $G_c$.  
{In fact, $\delta(g)$ encodes all the $\zz$ invariants for TCIs listed earlier, by choosing different operations $g$.  For instance, if $g$ is a translation then $\delta(g)$ is the corresponding $\zz$ weak invariant, if $g$ is inversion then $\delta(g)$ is the $\zz$ inversion invariant, and so on.} 
As an example, the topological crystal shown in Fig. \ref{fig:PC94}(b) has $\delta(\{1|100\})=0$, $\delta(\{1|010\})=0$, $\delta(\{1|001\})=1$, $\delta(\{4_{001}|\frac12\frac12\frac12\})=0$, $\delta(\{2_{110}|000\})=0$.
Taking advantage of the results of Ref.~\cite{Song2017a}, we find that these invariants, together with the mirror Chern number, uniquely label all the TCIs in our classification, and moreover we find all TCIs that are beyond layer construction by comparing with Ref.~[\onlinecite{Song2017a}] (Appendix~\ref{app:beyond}).

As suggested by the above discussion of invariants, the classification ${\cal C}$ of TCIs has a simple relationship with $H^1(G_c, \zz)$, which allows us to efficiently compute ${\cal C}$, and obtain the full TCI classification ${\cal C}$ for all space groups (Appendix~\ref{app:H1}).
%The prescription is to first compute $H^1(G_c, \zz)$, which is a product of $\zz$ factors, then replace $k$ of these factors with $\z$ factors to obtain ${\cal C}$, where $k$ is the number of symmetry-inequivalent sets of mirror planes in $G_c$. In fact, an equivalent procedure was shown in Ref.~\cite{Khalaf17symmetry} to give the classification of point group symmetric TCIs with non-trivial boundary modes; the same result was conjectured to hold for space group symmetry, a result we have now established. Using this method, we obtain the full TCI classification ${\cal C}$ for all space groups, as presented in Appendix~\ref{app:H1}.
Moreover, we find that there are 12 groups hosting topological crystals beyond layer construction; for such non-layer-constructable states, we tabulated their invariants and symmetry-based indicators \cite{Po2017,Bradlyn2017} in Appendix~\ref{app:beyond}, completing the mapping from indicators to TCI invariants \cite{Song2017a}.
Appendix~\ref{app:H1} also gives the classification of TCIs protected by point group symmetry for the 32 crystallographic point groups in three dimensions.

Finally, given the classification of TCIs, we obtained simple rules that extend this classification to include strong topological insulators (Appendix~\ref{app:unified}).   The key fact is that upon stacking two identical strong topological insulators together, one can either obtain a trivial state or a non-trivial TCI.  Upon identifying the state thus obtained, we obtain a a full classification of all topologically non-trivial insulators of non-interacting electrons with time reversal symmetry and significant spin-orbit coupling.

\section*{Discussion}

Our method provides a unified, real-space perspective for TCIs, complementary to the momentum-space perspective usually taken for free-fermion systems.  Real-space constructions have an advantage when electron interactions are considered.
For example, {because all $\zz$ TCIs are made from 2dTIs, from the stability of the 2dTI phase against interactions, we immediately know that the classification of $\zz$ TCIs is not affected by interactions.  %That is, the distinctions among $\zz$ TCIs are preserved in an interacting system.  
On the other hand, the $\z$ classification of $d=2$ MCI states collapses to $\z_8$ in the presence of interactions \cite{Morimoto2015}.  This implies that the classification of MTCIs does collapse, but the $\z$ invariants characterizing MTCIs are robust to interactions at least modulo eight.%(Further robust combinations of invariants may occur due to symmetries beyond a single mirror reflection).}

One can easily use the idea of topological crystals to classify free-electron $d=3$ insulators with time-reversal but without spin-orbit coupling; that is, with ${\rm SU}(2)$ spin rotation symmetry.  For such systems, time reversal and crystalline symmetry can be taken to act trivially on electron operators, with no factors of fermion parity, and in particular $T^2 = 1$.  It can then be seen that all 2-cells have symmetry class AI, while 1-cells can be in class AI or A.  In all these cases, only trivial states are possible, and no free-electron TCIs can occur.
%Therefore, there are simply no non-trivial ``building blocks'' for $d=3$ topological crystals, and we conclude that no free-electron TCIs can occur in this symmetry class.  
Interestingly, this means that any nonzero symmetry-based indicator implies some topological nodes in the bulk, proved by exhaustion in Ref. [\onlinecite{Song2017b}].

The topological crystal approach developed here can be applied in many other physical settings.  For instance, one can classify topological crystalline superconductors, described at the free-fermion level by Bogoliubov de Gennes Hamiltonians.  Moreover, as other works have begun to explore, it is also possible to use topological crystals to classify interacting fermionic cSPT phases.

%  In this case, the enumeration of topological crystals is expected to be richer because non-trivial  building blocks in both $d=1$ and $d=2$ will need to be considered, and because the results can depend on the superconducting pairing symmetry, which enters via the choice of $\omega_f \in H^2(G_c, \zz)$.  Finally, as other works have begun to explore in some simple cases \cite{song17topological, Watanabe2017}, it is possible to use topological crystals to classify interacting fermionic cSPT phases, including intrinsically interacting topological phases that require bulk interactions in order to exist.

\emph{Note added:}  As this work was being finalized for posting on the arXiv, Ref.~\cite{Shiozaki2018Generalized} appeared, which contains some related results.

\acknowledgments{MH is grateful to Ashvin Vishwanath for a useful discussion.  SZD and CF acknowledge support from  Ministry of Science and Technology of China under  Ministry of Science and Technology of China under grant numbers 2016YFA0302400,  2016YFA0300600, from National Science Foundation of China under grant number 11674370, and from Chinese Academy of Sciences under grant number XXH13506-202.  The research of SJH and MH is supported by the U.S. Department of Energy, Office of Science, Basic Energy Sciences (BES) under Award number DE-SC0014415.}

\bibliographystyle{apsrev4-1}
\bibliography{ref}

\clearpage
\newpage

\appendix

\onecolumngrid
\begin{center}
{\bf SUPPLEMENTARY MATERIAL}

\vspace{1cm}

\end{center}
\twocolumngrid

\section{Cell complex structure}
\label{app:cell_complex}

A cell complex is a topological space constructed by gluing together points (0-cells), and $n$-dimensional balls ($n$-cells).  In more detail, to construct a cell complex $X$, one starts with a set of discrete points $X^0$, referred to as the 0-skeleton.  Next one forms the 1-skeleton $X^1$ by attaching a set of 1-cells to $X^0$.  To attach a 1-cell, one starts with a closed interval on the real line (whose interior is the 1-cell), and the two endpoints are identified with points in $X^0$.  The process continues in the natural way; for instance, to attach a 2-cell to $X^1$, we start with a two-dimensional disc $D$ with boundary (whose interior is the 2-cell), and identify $\partial D$ with a subset of $X^1$ using a continuous map from $\partial D$ to $X^1$.  A more detailed discussion can be found in the book by Hatcher~\cite{hatcherbook}.

Here we describe in more detail how three-dimensional space ${\mathbb R}^3$ can be given a cell complex structure upon choosing an asymmetric unit (AU).  An AU is an open subset of ${\mathbb R}^3$ that is as large as possible, subject to the condition that no two points in the AU are related by the action of $G_c$.  The choice of AU is not unique.  While not strictly necessary, we can always choose an AU such that the boundary of the closure of the AU consists of segments of flat planes, \textit{i.e.} in the case of space group symmetry, the AU can be chosen as the interior of a polyhedron.  Once we choose an AU, it and its copies under the action of $G_c$ form the 3-cells, which are in one-to-one correspondence with elements of $G_c$.  The union of all the 3-cells is denoted by ${\cal A}$, and its complement $X^2 = {\mathbb R}^3 - {\cal A}$ is the 2-skeleton of the cell complex.

We choose 2-cells of $X^2$ satisfying three properties:  (1) Each 2-cell is a subset of a face where two 3-cells meet.  To be precise, we say that two 3-cells meet at a face when the intersection of their closures is homeomorphic to a 2-manifold (possibly with boundary), and we define the face where they meet to be this intersection.   
(2) No two \textit{distinct} points in the same 2-cell are related under the action of $G_c$.  Note that a  2-cell may be a subset of a mirror plane, in which case the mirror symmetry will take every point in the 2-cell to itself.  This property ensures that each 2-cell has no spatial symmetries; if there is a mirror symmetry, it acts on the 2-cell effectively as an internal symmetry.  (3)  The 2-cell structure on $X^2$ respects the $G_c$ symmetry. Precisely, given a 2-cell $e^2$ and a symmetry operation $g \in G_c$, the image $g(e^2)$ is also a 2-cell.

It is always possible to choose a set of 2-cells satisfying these properties:  we start with the set of faces where pairs of 3-cells meet, and take their interiors as 2-manifolds.  This gives a set of 2-cells for $X^2$ satisfying properties (1) and (3), but property (2) need not be satisfied.  This can be rectified by dividing up the 2-cells until property (2) is satisfied.

Letting ${\cal A}_2$ be the union of all the 2-cells, the 1-skeleton $X^1$ is $X^2 - {\cal A}_2$.  We choose 1-cells to satisfy three properties very similar to those for 2-cells.  A difference from the 2-cell case is that different numbers of 2-cells can meet at a 1-cell; we would like to ensure that the same set of $n$ 2-cells meets everywhere along the extent of a given 1-cell.  We therefore modify property (1) as follows:  each 1-cell is a subset of an edge where exactly $n$ 2-cells meet.  More precisely, we say that $n$ 2-cells meet at an edge when the intersection of their closures is homeomorphic to a 1-manifold (possibly with boundary), and the edge where they meet is defined to be this intersection.  Apart from these $n$ 2-cells, we require the edge to have empty intersection with the closure of any other 2-cell.
 Properties (2) and (3) are required to hold with the obvious modifications.

The 0-cells are just the points where two or more 1-cells meet.  Formally, letting ${\cal A}_1$ be the union of all 1-cells, the 0-cells are the points of $X^1 - {\cal A}_1$.

We  illustrate this rather abstract discussion with some examples.  First, we consider $G_c = C_i$, the point group generated by inversion symmetry.  We take the AU to be the half space $z >0$, and the 3-cells are then the two half-spaces $z >0$ and $z < 0$.  There are two 2-cells, which are $z = 0$ half planes with $y > 0$ and $y < 0$, and two 1-cells, which are $z = y = 0$ half lines with $x >0$ and $x < 0$.  Finally, the single 0-cell is the point at the origin.

As a second example, we take $G_c$ to be space group \#1, which consists only of translation symmetry.  We set the lattice constant to unity and take the three Bravais lattice basis vectors to be $(1,0,0)$, $(0,1,0)$ and $(0,0,1)$.  A natural choice for an AU is simply the interior of a unit cell, \textit{i.e.} the region $0 < x,y,z < 1$.  The 3-cells are then the copies of the AU under translation.  There are three kinds of 2-cells.  One type consists of the $xy$ plane (\textit{i.e.} $z=0$ plane) region with $0 < x,y < 1$ and its images under translation, and the other two types are similar but lie in $xz$ and $yz$ planes.  Similarly, there are three kinds of 1-cells, with one type consisting of the $x = y = 0$ region with $0 < z < 1$ and its images under translation.  The other two types are similar regions oriented along the $x$ and $y$ axes.  Finally, the 0-cells are points $(n_x, n_y, n_z)$, with $n_x, n_y, n_z$ integers.

\section{Relativistic Dirac Hamiltonian and symmetry action on electrons}
\label{app:relativistic}

Here, we sketch how the action of symmetry on electrons with significant spin-orbit coupling and time reversal symmetry can be determined from the relativistic Dirac Hamiltonian describing electrons in vacuum.  By ``action of symmetry,'' we are referring to the fact that certain equations in the symmetry group can be modified by factors of $(-1)^F$ (fermion parity), and these factors can be determined by studying the Dirac Hamiltonian.  More formally, this corresponds to determining an element $\omega_f \in H^2(G, \zz)$ that specifies how a symmetry group $G$ acts on fermion fields.

The Dirac Hamiltonian is
\begin{equation}
H_D = \int d^3 \br \, \Psi^\dagger(\br) \Big[ \vec{\alpha} \cdot (- i \vec{\nabla} ) + \beta m \Big] \Psi(\br) \text{,}
\end{equation}
where $\Psi(\br)$ is the four-component Dirac field,
\begin{equation}
\alpha^i = \left( \begin{array}{cc} 
0 & \mu^i  \\ \mu^i & 0 
\end{array}\right) = \mu^i \tau^1 \text{,}
\end{equation}
and
\begin{equation}
\beta = \left(\begin{array}{cc} 
\mathbbm{1} & 0 \\
0 & - \mathbbm{1} 
\end{array}\right) = \tau^3 \text{,}
\end{equation}
with $2 \times 2$ Pauli matrices $\mu^i$ and $\tau^i$.

$H_D$ is invariant under the group of rigid motions of three-dimensional space and time reversal symmetry.
\begin{equation}
T : \Psi \to (i \mu^2) \Psi \text{.}
\end{equation}
An arbitrary rigid motion can be obtained by composing translations, rotation, and inversion.  Inversion acts by
\begin{equation}
{\cal I} : \Psi(\br) \to \tau^3 \Psi(-\br) \text{,}
\end{equation}
and translation by $\boldsymbol{a}$ acts by $\Psi(\br) \to \Psi(\br + \boldsymbol{a})$.  Finally, we consider a rotation $R$ by an angle $\theta$ about an axis $\hat{n}$, which acts by
\begin{equation}
R : \Psi(\br) \to \exp\Big(\frac{i}{2} \vec{\theta} \cdot \vec{\mu} \Big)   \Psi(R \br) \text{,}
\end{equation}
where $\vec{\theta} = \theta \hat{n}$.  We can see from these results that $T^2 = (-1)^F$, and that $T g = g T$ for $g$ any rigid motion.

If we add a scalar potential with the symmetry of the crystal lattice, then the continuum symmetry is broken to the desired crystalline symmetry $G_c$.  Given an equation that holds in the group $G_c$, the action on the Dirac field of each rigid motion appearing in the equation is determined from the results above.  Most importantly, the Dirac matrix structure is determined, and from this matrix structure we can determine whether or not the equation is modified by a factor of $(-1)^F$.  For example, the mirror reflection $\sigma : z \to -z$ acts on the Dirac field by
\begin{equation}
\sigma : \Psi(\br) \to i \tau^3 \sigma^3 \Psi(\br') \text{,} \label{eqn:sigma-Dirac}
\end{equation}
and therefore $\sigma^2 = (-1)^F$.

\section{Formal structure of classification resolved by block dimension}
\label{app:formal}

Here we describe the Abelian group structure of TCIs, and how taking a certain quotient allows us to ignore distinctions among atomic insulators.  We let ${\cal D}_{d_b}$ be the Abelian group classifying insulators whose non-trivial building blocks are dimension $d_b$ and below.    That is, states classified by ${\cal D}_{d_b}$ can be reduced to a state on $X^2$ where all $n$-cells with $n > d_b$ host a trivial state.  Clearly, $d_b = 0,1,2$, and we have the sequence of subgroups ${\cal D}_0 \subset {\cal D}_1 \subset {\cal D}_2$. ${\cal D}_2$ is the classification of all TCIs, or at least all those that can be classified in terms of topological crystals.  The observation that all 1-cells are trivial implies ${\cal D}_0 = {\cal D}_1$.  Phases in ${\cal D}_0$ are atomic insulators, which we wish to exclude from consideration.  Even though there are distinct atomic insulators constituting different quantum phases of matter, all atomic insulators are in some sense topologically trivial.  We can eliminate these states by taking the quotient  ${\cal C} = {\cal D}_2 / {\cal D}_0$, which gives the desired classification of TCIs.

\section{Gluing MCI building blocks: planar decomposition of mirror TCIs}
\label{app:MCIgluing}

Here we address the consequences of the gluing conditions for MTCIs, \textit{i.e.} topological crystals built from MCI building blocks placed on the 2-cells of $X^2$.  In particular, we show that MTCIs can always be decomposed into decoupled planar MCI states placed on mirror planes.  We consider a mirror plane $P$, and note that we must have $P \subset X^2$.  Therefore, up to a set of measure zero, $P$ is a union of 2-cells of $X^2$.  We consider a 2-cell $e^2_1 \subset P$, and place a MCI state on $e^2_1$ whose invariant is some element of $\z$.  We want to show that symmetry and gluing along 1-cells implies that every 2-cell in $P$ is a MCI with the same $\z$ invariant as the state in $e^2_1$.  It is enough to show that this holds for a single 2-cell $e^2_2 \subset P$ that is adjacent to $e^2_1$ in $P$.  That is, $e^2_1$ and $e^2_2$ meet at a 1-cell $e^1 \subset P$, as illustrated in Fig.~\ref{fig:e1e2}.

\begin{figure}[h]
\includegraphics[width=.2\textwidth]{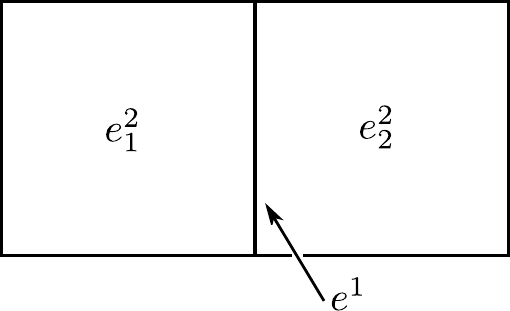}
\caption{Illustration of the 2-cells $e^2_1$ and $e^2_2$ and the 1-cell $e^1$ used to discuss the effect of gluing conditions on MTCIs.
\label{fig:e1e2}}
\end{figure}

To proceed, we consider a number of cases.  In case (1), there exists an element $g \in G_c$ that maps $e^2_1$ to $e^2_2$.  First, we show that symmetry requires that both 2-cells host an MCI state with the same invariant (this is not \textit{a priori} obvious; it is conceivable that some symmetry operations could change the sign of the invariant).  To begin, we claim that either $g \sigma = \sigma g $, when $g$ preserves the orientation of the mirror plane, or $g \sigma = (-1)^F \sigma g$, when $g$ reverses the orientation of the mirror plane.  If we ignore factors of fermion parity, then $g \sigma = \sigma g$, or equivalently $g \sigma g^{-1} = \sigma$.  To see this, we observe that $g \in G_P$, where $G_P \subset G_c$ is the group of symmetries of the mirror plane. Moreover, $\sigma$ is the only non-trivial element of $G_P$ that acts on the mirror plane as the identity rigid motion.  The operation $g \sigma g^{-1}$ also acts on the mirror plane as the identity rigid motion, and $\sigma$ cannot be conjugate to the identity in $G_P$, therefore $g \sigma g^{-1} = \sigma$.

The relativistic Dirac Hamiltonian as discussed in Appendix~\ref{app:relativistic} allows us to determine the presence or absence of the $(-1)^F$ factor. We choose coordinates so that the mirror plane is the $z=0$ plane, and the action of $\sigma$ on the Dirac field $\Psi(\br)$ is given in Eq.~(\ref{eqn:sigma-Dirac}).  We consider a symmetry operation $g$ that takes the mirror plane into itself, with action on the Dirac field
\begin{equation}
g : \Psi(\br) \to M_g \Psi(\br') \text{,}
\end{equation}
where
\begin{equation}
\br' = O \br + \vec{t} \text{,}
\end{equation}
where $O$ is an orthogonal matrix.   The requirement that the mirror plane goes into itself under $g$ implies that $t_z = O_{zx} = O_{zy} = 0$.  Moreover, because $O$ is an orthogonal matrix, $O_{xz} = O_{yz} = 0$, and $O_{zz} = \pm 1$.  We are free to multiply $g$ by inversion and/or translations within the $z=0$ plane to make $g$ into a rotation.  This can be done because both translations and inversion preserve the orientation of the $z=0$ plane.  Moreover, translations have no effect on $M_g$, while inversion commutes with $\sigma$.  After doing this, there are two possibilities for $g$.  One possibility is a rotation by $\theta$ with axis normal to the plane; this operation preserves the orientation of the plane, and we have $M_g = \exp(i \theta \mu^3 / 2)$, so that $g$ commutes with $\sigma$.  The other possibility is a $C_2$ rotation with axis normal to the plane; this operation reverses the orientation of the plane and anti-commutes with $\sigma$.  This establishes the claim that $g \sigma = \sigma g$, when $g$ preserves the orientation of the mirror plane, or $g \sigma = (-1)^F \sigma g$, when $g$ reverses the orientation of the mirror plane

Now we employ this claim to show that the MCI states on $e^2_1$ and $e^2_2$ have the same $\z$ invariant.  Consider a one-electron state $| \psi \rangle$ supported only on $e^2_1$, whose mirror eigenvalue is given by $\sigma | \psi \rangle = i | \psi \rangle$.  The state $g | \psi \rangle$ is supported on $e^2_2$, and has mirror eigenvalue $+i$ if $g$ preserves the orientation of the plane, and $-i$ if it reverses orientation of the plane.  Because the Chern number of each sector with fixed mirror eigenvalue is preserved when $g$ preserves orientation, and reversed when $g$ reverses orientation, it follows that the two MCI states have the same $\z$ index.

To complete the discussion of case (1), we need to address gluing of the two MCI states at $e_1$.  It is enough to consider only symmetries that take the set of cells $\{ e^1, e^2_1, e^2_2 \}$ into itself.  There are two sub-cases.  In case (1a), the only relevant symmetry is the mirror reflection itself.  In this case it is obvious that two MCI states with the same invariant can be glued together along $e^1$.  In case (1b),  $e^1$ is contained within a $C_{2 v}$ axis.  To analyze gluing at $e^1$, we study the edge theory at $e^1$ for MCI states on $e^2_1$ and $e^2_2$.  The edge of $e^2_1$ ($e^2_2$) consists of a pair of counter-propagating fermion modes $c_{R1}$ and $c_{L1}$ ($c_{R2}$ and $c_{L2}$).  We assemble these 1d fermions into the four-component field $\psi^T = \left( \begin{array}{cccc} c_{R1} & c_{L1} & c_{R2} & c_{L2} \end{array} \right)$.  Denoting by $\sigma'$ the mirror symmetry exchanging the two 2-cells, we take the symmetries to act by
\begin{eqnarray} 
T : \psi &\to& (i \mu^2) \psi \\
\sigma : \psi &\to& i \mu^3 \tau^3 \psi \\
\sigma' : \psi &\to& i \mu^3 \tau^1 \psi \text{,} \label{eqn:sigmaprime}
\end{eqnarray}
where the $\mu^i$ and $\tau^i$ Pauli matrices act in the $4 \times 4$ matrix space just as in the earlier discussion of the relativistic Dirac Hamiltonian.  These symmetries act appropriately on fermions, and are compatible with the two 2-cells having the same MCI index.  These symmetries allow the mass term $\psi^\dagger \mu^2 \tau^2 \psi$, which gaps out the fermions on $e^1$ and thus glues the two 2-cells together.

Now we move on to case (2), where there is no element $g \in G_c$ that maps $e^2_1$ into $e^2_2$.  In this case, symmetry does not determine which state is placed on $e^2_2$, but we will see that this is determined by gluing at $e^1$.  We will find it useful to consider the 3-cells that meet at $e^2_1$ and $e^2_2$.  These are defined in Fig.~\ref{fig:3cells}.  We consider two sub-cases.  In case (2a), $e^2_{1 \uparrow} = e^2_{2 \uparrow}$.  It follows immediately that $e^2_{1 \downarrow} = e^2_{2 \downarrow}$, so there is only a single 3-cell above and below the mirror plane in the region shown in Fig.~\ref{fig:3cells}.  This means that $e^2_1$ and $e^2_2$ are the only 2-cells meeting at $e^1$, which further implies that the only symmetry taking $e^1$ into itself is the mirror symmetry.  The the gluing condition at $e^1$ is then satisfied if and only if an MCI state is placed in $e^2_2$.

\begin{figure}[h]
\includegraphics[width=.35\textwidth]{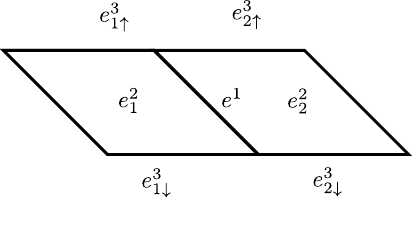}
\caption{The 3-cells $e^3_{1\uparrow}$ and $e^3_{1\downarrow}$ ($e^3_{2\uparrow}$ and $e^3_{2\downarrow}$) meet at the 2-cell $e^2_1$ ($e^2_2$).  
\label{fig:3cells}}
\end{figure}

In case (2b), $e^2_{1 \uparrow} \neq e^2_{2 \uparrow}$, which implies that $e^2_1$ and $e^2_2$ are not the only 2-cells meeting at $e^1$.  The additional 2-cells come in mirror-symmetric pairs above and below the mirror plane.  There are two further sub-cases.  In case (2b.i), the only symmetry taking $e^1$ into itself is the mirror symmetry.  In this case, the additional pairs of 2-cells do not coincide with mirror planes.  Therefore, for each pair the only non-trivial possibility is that both 2-cells host a 2dTI state, in which case the 2dTI edges of the pair can be gapped out at $e^1$.  The relevant edge theory for each pair at $e^1$  is the same as that discussed in case (1), except with only $\sigma'$ and $T$ symmetry (\textit{i.e.} without  $\sigma$ symmetry), and it follows from the discussion there that this edge can be gapped.  Therefore, these additional pairs of 2-cells can be effectively eliminated, and the gluing condition again requires us to place a MCI state on $e^2_2$.

\begin{figure}[h]
\includegraphics[width=.3\textwidth]{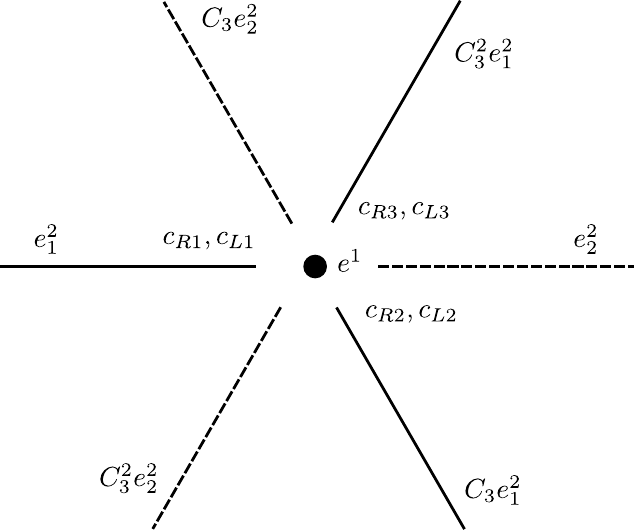}
\caption{Cross section through $e^1$ and the 2-cells that meet at $e^1$, in case (2b.ii), where $e^1$ is contained in a $C_{3v}$ axis.  In the text we consider placing an MCI state on $e^2_1$ and its rotation images (solid lines), while placing a trivial state on $e^2_2$ and its rotation images (dashed lines), and show that the gluing conditions at $e^1$ cannot be satisfied.  The locations of the 1d fermion modes ($c_{Ri}, c_{Li}$) at the $e^1$ edge of each MCI state are shown.
\label{fig:c3v}}
\end{figure}

Finally, in case (2b.ii), $e^1$ is contained in a $C_{3v}$ axis, where the $C_{3v}$ symmetry is generated by $\sigma$ and a 3-fold rotation $C_3$.  Here, there are six 2-cells that coincide with the  mirror planes meeting at $e^1$.  These 2-cells come in three pairs, with each pair contained in one of the three mirror planes that intersect $e^1$.  $e^2_1$ and $e^2_2$ constitute one such pair, with the other two pairs obtained from it under the 3-fold rotation.  We suppose that a MCI state is placed on $e^2_1$ and its rotation images, but not on $e^2_2$ (see Fig.~\ref{fig:c3v}); we will show that it is impossible to gap out the resulting edge theory at $e^1$, which will imply that, again, a MCI state must be placed on $e^2_2$.  The edge fermions for the $e^2_1$ MCI are $c_{R1}$ and $c_{L1}$, with
\begin{equation}
\sigma : \left\{ \begin{array}{l}
c_{R1} \to i c_{R 1} \\
c_{L1} \to -i c_{L 1} 
\end{array}\right. \text{.}
\end{equation}
The images of these fermions under $C_3$ rotation are $c_{R2} = C_3 c_{R1} C^{-1}_3$ and $c_{R3} = C^2_3 c_{R1} C^{-2}_3$, with identical expressions holding for the left-moving modes.  The generators of the $C_{3v}$ group satisfy the following relations, acting on a fermion field:
\begin{eqnarray}
\sigma^2 &=& -1 \\
C_3^3 &=& -1 \\
(\sigma C_3)^2 &=& -1 \text{.}
\end{eqnarray}
Using these relations, we find
\begin{eqnarray}
\sigma  : \left\{ \begin{array}{l}
c_{R 2} \to -i c_{R 3} \\
c_{L 2} \to i c_{L 3} \\
c_{R 3} \to -i c_{R 2} \\
c_{L 3} \to i c_{L 2} 
\end{array}\right. \text{.}
\end{eqnarray}
Now focusing only on the mirror reflection symmetry, we can change variables to diagonalize $\sigma$ and find that the $c_{R2}, c_{R3}, c_{L2}, c_{L3}$ fermion modes can be gapped out.  This leaves the $c_{R1}$, $c_{L1}$ edge of the MCI state on $e^2_1$, which cannot be gapped; this establishes the desired result.

\section{Gluing condition for $\zz$ TCIs}
\label{app:Z2gluing}

Here, we consider the gluing condition for $\zz$ TCIs, \emph{i.e.} the requirement that there are no gapless modes within the bulk.  If we consider placing 2dTI states on a subset of the 2-cells of $X^2$, we will show that the gluing condition can be satisfied if and only if an even number of 2dTI edges meet at each 1-cell.  This is illustrated in Fig.~\ref{fig:gluing} for the space groups $P3$ (no. 143) and $P4$ (no. 75).  One direction is trivial to show:  if the gluing condition is satisfied, then an even number of 2dTI edges must meet at each 1-cell $e^1$, because otherwise time reversal would forbid $e^1$ from being gapped.

\begin{figure}
\begin{centering}
\includegraphics[width=0.9\linewidth]{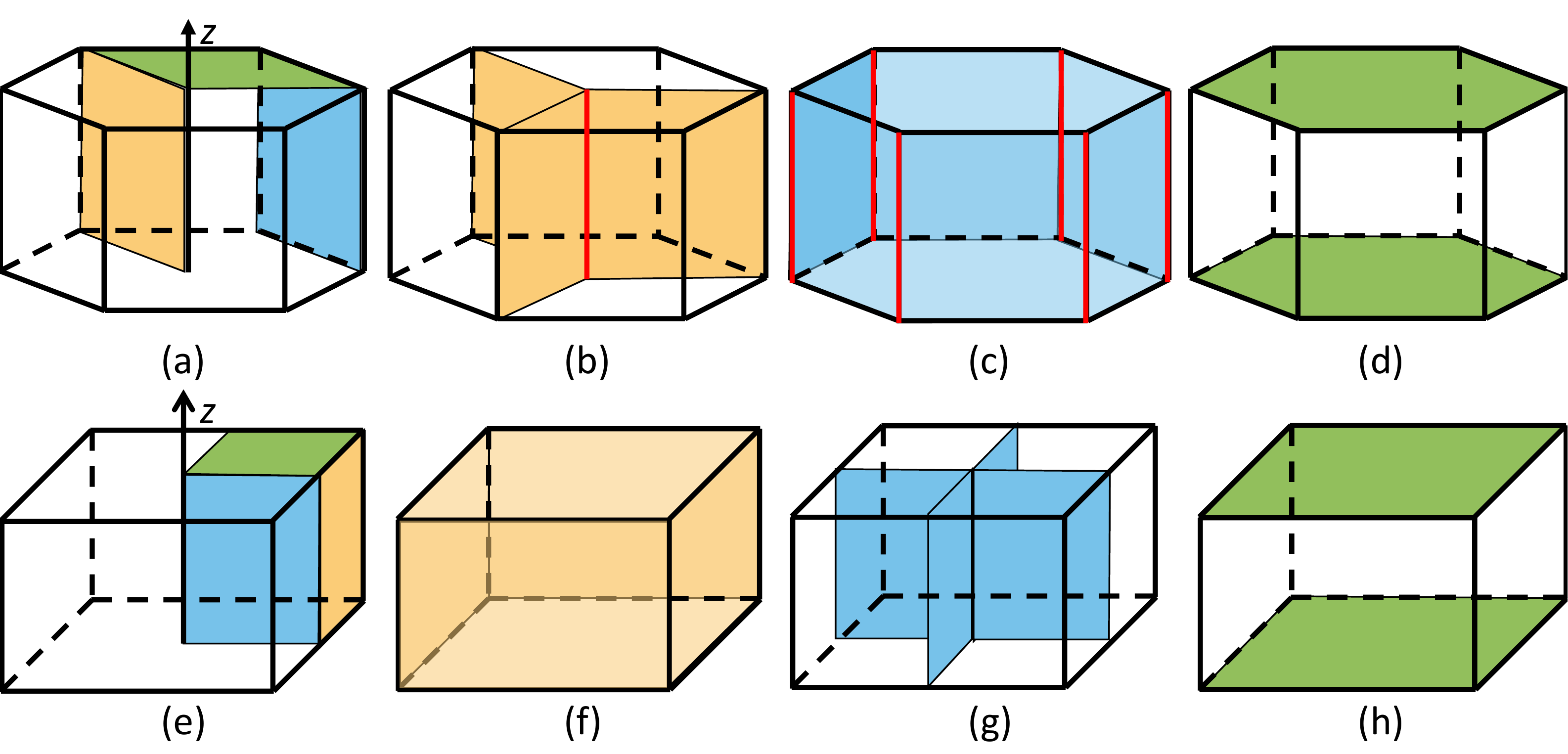}
\par\end{centering}
\caption{Illustration of the effects of the gluing condition for $\zz$ TCIs for two representative space groups.  Panels (a) through (d) show $\zz$ TCIs for the space group $P3$ (no. 143).  (a) shows the three symmetry-inequivalent 2-cells.  (b) and (c) show putative $\zz$ TCIs obtained by placing 2dTI states on the orange and blue 2-cells respectively.  These states are forbidden by the gluing condition; they each have three 2-cells of 2dTI meeting at a three-fold axis (red line), where time-reversal symmetry forbids gapping of the odd number of edge modes.  (d) shows an allowed $\zz$.  Panels (e) through (h) show $\zz$ TCIs for the space group $P4$ (no. 75), with (e) again showing the three symmetry-inequivalent 2-cells.  For this space group, the three $\zz$ TCIs obtained by placing 2dTIs on each type of 2-cell are shown in panels (f) through (h) and all are allowed by the gluing condition.  States (f) and (g) both have four 2dTI 2-cells meeting at four-fold axes, where, as discussed in the text, the edge modes can be gapped.}\label{fig:gluing}
\end{figure}

Now we suppose that we place 2dTI states on 2-cells of $X^2$ such that an even number of 2dTI edges meet at each 1-cell.  We would like to show that each 1-cell can be gapped, and thus the gluing condition is satisfied.  We do this by considering the different possible point group symmetries of a 1-cell $e^1$, which may impose constraints on gluing of 2dTI edge modes along $e^1$.  If $e^1$ has trivial point group symmetry, then the only symmetries are charge conservation and time reversal, and an even number of 2dTI edge modes can always be gapped.  If $e^1$ is contained in a mirror plane, 2dTI edge modes come in mirror-symmetric pairs above and below the mirror plane, and we have already shown in Appendix~\ref{app:MCIgluing} that each such pair can be gapped.

Next, we consider the case where $e^1$ is contained in a $C_n$ axis. When $n = 2$, 2dTI edge modes come in pairs related by $C_2$ symmetry, and the action of $C_2$ symmetry on such a pair is identical to that of the $\sigma'$ symmetry discussed in Appendix~\ref{app:MCIgluing} (see Eq.~\ref{eqn:sigmaprime} and surrounding discussion).  Therefore such pairs of edge modes can be gapped.  When $n > 2$ is even, edge modes come in groups of $n$ related by $C_n$ symmetry, and these can be grouped into $n/2$ pairs related by $C_2$ symmetry.  We can focus on one such pair and gap it out, then use the $C_n$ symmetry to ``copy'' its mass term to the other $n/2 -1$ pairs, which gaps out all the modes respecting the $C_n$ symmetry.  Finally, for $C_3$ symmetry, edge modes must come in groups of six, with two sets of three edge modes related by symmetry.  We can take a pair of edge modes unrelated by symmetry and gap these out, then use the $C_3$ symmetry to copy the resulting mass term to the other two pairs of edge modes.

Similar arguments can be applied when $e^1$ is contained in a $C_{nv}$ axis.  Here,  2-cells that can host 2dTIs lie away from the mirror planes, so that edge modes come in groups of $2n$ related by $C_{nv}$ symmetry.  Viewing $C_{nv}$ as generated by a mirror reflection and $C_n$ rotation, we can start by gapping out a pair of neighboring edge modes related by the mirror symmetry, then copying its mass term using the rotational symmetry.

\section{Gluing at 0-cells}
\label{app:0gluing}

In the analysis of the gluing condition in Appendices~\ref{app:MCIgluing} and~\ref{app:Z2gluing}, we started with a set of topological states on 2-cells, and considered gluing these states together at 1-cells.  In principle, this may not be the end of the story; we need to consider gluing at 0-cells.  That is, we need to ensure there are no localized protected gapless states at 0-cells, which would violate the gluing condition.  However, there is a simple reason this cannot occur:  consider the set of gapped 1-cells that meet at a 0-cell.  We can lump these 1-cells together and view them as a single gapped $d=1$ system, with the 0-cell as its endpoint.  Upon decomposing the spectrum into irreducible representations of any site symmetry at the 0-cell, this $d=1$ system divides into sectors that are either in class A or AII.  In order to have a protected gapless state on the 0-cell, the $d=1$ system would need to be topologically non-trivial, but class A and AII have a trivial classification in $d=1$.

\section{Topological crystals, topological invariants, and $H^1(G,\zz)$}
\label{app:H1}

In this section we give a more detailed discussion of the $\zz$-valued function of invariants $\delta(g)$ characterizing a topological crystal.  Based on this discussion, we establish a connection between the classification ${\cal C}$ of TCIs and $H^1(G_c, \zz)$, which allows ${\cal C}$ to be computed with the aid of standard computer algebra tools such as GAP \cite{GAP4}.  In particular, we show that ${\cal C}$ and $H^1(G_c, \zz)$ have the same number of generators.

In the text we defined $\delta(g)$ only for a $\zz$ TCI.  In fact, it will be useful for our present purposes to give a definition valid for an arbitrary topological crystal.  First, we introduce the notion of a $\zz$ coloring of $X^2$, which is given by associating a $\zz$ number to each 2-cell of $X^2$, so that each 2-cell is either colored and assigned 1, or empty and assigned 0.  These $\zz$ numbers must be assigned to respect the crystalline symmetry, and satisfy a gluing condition, namely for each 1-cell, an even number of the 2-cells meeting there must be colored.  $\zz$ colorings can be added using the $\zz$ addition law, and this makes $\zz$ colorings into a group that we denote by $\tC$.   (We remark that $\zz$ colorings can be viewed as elements of the homology group $H_2(X^2; \zz)$ satisfying a symmetry condition, but we will not make use of this here.)

Next, we observe that there is a map from topological crystals (elements of ${\cal C}$) to $\tC$.  We denote this map by $\pi : {\cal C} \to \tC$.  Empty cells of the topological crystal map to empty cells in the $\zz$ coloring.  Cells decorated with 2dTI map to colored cells.  Cells decorated with a MCI state map to colored cells when the mirror Chern number is odd, and to empty cells when it is even.  (We are using the convention that the smallest possible mirror Chern number is 1; some authors use a definition of mirror Chern number that is twice our definition.)  

Finally, given a $\zz$ coloring of $X^2$, we define $\delta(g)$ as in the main text.  That is, we arbitrarily choose one AU, let ${\bf r}$ be a point inside, and define $\delta(g) =1$ if a path connecting ${\bf r}$ to $g {\bf r}$ crosses an odd number of colored 2-cells, while $\delta(g) = 0$ if the path crosses and even number of colored 2-cells.  The path should be chosen to avoid 0-cells and 1-cells, but is otherwise an arbitrary continuous path.

We now establish some properties of $\delta(g)$ quoted in the main text.  First, we show that $\delta(g)$ is independent of the path chosen to connect ${\bf r}$ to $g {\bf r}$, and is thus a well-defined function mapping $G_c$ to $\zz$.  Any two such paths are related by a finite number of moves, where a segment of the path is passed through a 1-cell.  The gluing condition says that an even number of colored 2-cells meet at every 1-cell, so such moves do not affect $\delta(g)$.  At this stage, we have not yet shown that $\delta$ is independent of the arbitrary choice of AU.

Next, we show that $\delta$ is in fact a homomorphism from $G_c$ to $\zz$, that is $\delta(g_1 g_2) = \delta(g_1) + \delta(g_2)$.  We compute $\delta(g_1 g_2)$ by considering a path from ${\bf r}$ to $g_1 g_2 {\bf r}$ that first goes from ${\bf r}$ to $g_1 {\bf r}$, and then goes from $g_1 {\bf r}$ to $g_1 g_2 {\bf r}$.  The number of colored 2-cells modulo two crossed by the first segment is $\delta(g_1)$, by definition.  The second segment is related by symmetry to a path joining ${\bf r}$ to $g_2 {\bf r}$, and so $\delta(g_2)$ is the number of colored 2-cells (modulo two) crossed by the second segment.  Therefore $\delta(g_1 g_2) = \delta(g_1) + \delta(g_2)$.

Finally, we show that $\delta$ does not depend on the arbitrary choice of AU.  Let $\delta(g)$ be the function defined by choosing an AU with a point ${\bf r}$ inside, and let $\delta'(g)$ be the function defined by choosing a different AU, which contains a point $g_0 {\bf r}$ for some $g_0 \in G_c$.  Then $\delta'(g)$ is the number of colored 2-cells (modulo two) crossed by a path connecting $g_0 {\bf r}$ to $g g_0 {\bf r}$.  By symmetry, this number is the same as for a path joining ${\bf  r}$ to $g_0^{-1} g g_0 {\bf r}$, which shows that $\delta'(g) = \delta(g_0^{-1} g g_0)$.  But this implies $\delta'(g) = \delta(g)$, because $\delta$ is a homomorphism and $\zz$ is Abelian.

Our construction of $\delta$ gives a map $\Delta : \tC \to H^1(G_c, \zz)$, where $H^1(G_c, \zz)$ is viewed as the group of homomorphisms from $G_c$ to $\zz$.  In fact, $\Delta$ is an isomorphism, so  $\tC \simeq H^1(G_c, \zz)$. It is easy to see that $\Delta$ is injective.  To see that $\Delta$ is surjective, we need to show that given $\delta : G_c \to \zz$, we can construct a corresponding $\zz$ coloring.  Upon arbitrarily choosing an AU, the 3-cells of ${\mathbb R}^3$ are in one-to-one correspondence with elements of $G_c$.  We then color each 3-cell with the $\zz$ number $\delta(g)$.  Given a 2-cell, let $g_1$ and $g_2$ label the two 3-cells that meet at the 2-cell.  We then color the 2-cell with the $\zz$ number $\delta(g_1) + \delta(g_2)$.  The resulting assignment of $\zz$ numbers to 2-cells is clearly symmetric and satisfies the gluing condition, and is thus a $\zz$ coloring of $X^2$.  By construction, $\Delta$ maps this $\zz$ coloring to $\delta$.

Now that we have shown $\tC \simeq H^1(G_c, \zz)$, we would like to show that ${\cal C}$ and $\tC$ have the same number of generators.  First, we observe that ${\cal C} = {\cal C}_{{\rm MCI}} \times {\cal C}_{{\rm 2dTI}}$, where ${\cal C}_{{\rm MCI}}$ is the classification of MTCIs (\emph{i.e.}, topological crystals built from MCI states) and ${\cal C}_{{\rm 2dTI}}$ is the classification of $\zz$ TCIs.  ${\cal C}_{{\rm MCI}}$ is a product of $\z$ factors, and ${\cal C}_{{\rm 2dTI}}$ is a product of $\zz$ factors.    We introduce a similar decomposition $\tC = \tC_{{\rm MCI}} \times \tC_{{\rm 2dTI}}$, where $\tC_{{\rm MCI}}$ is defined to be the subgroup of $\zz$ colorings whose colored 2-cells lie in mirror planes, and $\tC_{{\rm 2dTI}}$ is the subgroup of $\zz$ colorings where all 2-cells lying in mirror planes are empty.  Clearly $\tC$, $\tC_{{\rm MCI}}$ and $\tC_{{\rm 2dTI}}$ are all products of $\zz$ factors.    To prove that $\tC = \tC_{{\rm MCI}} \times \tC_{{\rm 2dTI}}$, it is enough to show that an arbitrary $\zz$ coloring $c \in \tC$ can be written uniquely as $c = c_m c_{\bar{m}}$ for some $c_m \in \tC_{{\rm MCI}}$ and $c_{\bar{m}} \in \tC_{{\rm 2dTI}}$.  Given $c \in \tC$, we consider a 1-cell $e^1$ contained in a mirror plane.  $n$ 2-cells meet at $e^1$, two of which lie in the mirror plane, and $n - 2$ of which lie outside the mirror plane.  The $n-2$ 2-cells outside the mirror plane can be grouped into pairs related by mirror reflection, so that the two 2-cells in each pair are either both colored or both empty.  It follows that the two 2-cells contained in the mirror plane are also either both colored or both empty.  Therefore, we can define a new $\zz$ coloring $c_m \in \tC_{{\rm MCI}}$ by starting with $c$ and replacing all colored 2-cells not lying in mirror planes with empty cells.  Similarly, if we replace all the colored 2-cells within mirror planes with empty cells, we obtain $c_{\bar{m}} \in \tC_{{\rm 2dTI}}$.  It is obvious that $c = c_m c_{\bar{m}}$, and that $c_m$ and $c_{\bar{m}}$ are unique.

Using the above discussion, we can show that the map $\pi : {\cal C} \to \tC$ gives a one-to-one correspondence between generators of ${\cal C}$ and $\tC$, so the two groups have the same number of generators.  First, restricting $\pi$ to ${\cal C}_{{\rm 2dTI}}$ gives an isomorphism between ${\cal C}_{{\rm 2dTI}}$ and $\tC_{{\rm 2dTI}}$, so these subgroups clearly have the same number of generators.  Second, we can take each $\z$ factor of ${\cal C}_{{\rm MCI}}$ to be generated a topological crystal obtained by decorating the 2-cells of a mirror plane, as well as all symmetry-equivalent mirror planes, with a MCI state of unit Chern number.  The above discussion implies that $\tC_{{\rm MCI}}$ is generated by $\zz$ colorings obtained by coloring all the 2-cells of set of symmetry-equivalent mirror planes, and these generators are images of the ${\cal C}_{{\rm MCI}}$ generators under $\pi$, giving a one-to-one correspondence between generators of ${\cal C}_{{\rm MCI}}$ and $\tC_{{\rm 2dTI}}$.

These results make it a simple matter to compute the TCI classification ${\cal C}$.  First, we compute $H^1(G_c, \zz)$; this can be done using GAP \cite{GAP4}.  Then, we know that the number of $\z$ factors in ${\cal C}$ is $n_M$, the number of symmetry-inequivalent sets of mirror planes.  We then obtain ${\cal C}$ from $H^1(G_c, \zz)$ by replacing $n_M$ of the $\zz$ factors with $\z$ factors.  For $G_c$ a crystallographic point group or space group, $n_M$ can be obtained immediately from information tabulated in the International Tables for Crystallography \cite{inttables}.  The results of this procedure are presented in Table~\ref{tab:PG} and crystalline point groups, and in Table~\ref{tab:SG} for space groups.  As discussed in the main text, we note that Khalaf \emph{et. al.} have also obtained the results in these tables via a mathematically equivalent procedure \cite{Khalaf17symmetry}.

While useful, we emphasize that this largely automated procedure is not a substitute for explicit real-space construction of topological crystals for a given symmetry group of interest, as described in the main text.  The latter procedure results in the same group structure, but also provides additional physical insight and a starting point for further analysis, by giving an explicit real-space construction of each of the TCI phases classified.  In fact, we also obtained the results in Table~\ref{tab:SG} by automating the explicit real-space constructions.

\renewcommand\arraystretch{1.15}
\begin{table}
\begin{centering}
\begin{tabular}{|c|c|}
\hline
Point group &  TCI Classification  \\
\hline
$1$ &  N/A  \\
\hline
$\overline{1}$ &  $\zz$  \\
\hline
$2$ &  $\zz$  \\
\hline
$m$ &  $\z$  \\
\hline
$2/m$ &  $\z \times \zz$  \\
\hline
$222$ &  $\zz^{2}$  \\
\hline
$mm2$ &  $\z^{2}$  \\
\hline
$mmm$ &  $\z^{3}$  \\
\hline
$4$ &  $\zz$  \\
\hline
$\overline{4}$ &  $\zz$  \\
\hline
$4/m$ &  $\z \times \zz$  \\
\hline
$422$ &  $\zz^{2}$  \\
\hline
$4mm$ &  $\z^{2}$  \\
\hline
$\overline{4}2m$ &  $\z \times \zz$  \\
\hline
$4/mmm$ &  $\z^{3}$  \\
\hline
$3$ &  N/A  \\
\hline
$\overline{3}$ &  $\zz$  \\
\hline
$32$ &  $\zz$  \\
\hline
$3m$ &  $\z$  \\
\hline
$\overline{3}m$ &  $\z \times \zz$  \\
\hline
$6$ &  $\zz$  \\
\hline
$\overline{6}$ &  $\z$  \\
\hline
$6/m$ &  $\z \times \zz$  \\
\hline
$622$ &  $\zz^{2}$  \\
\hline
$6mm$ &  $\z^{2}$  \\
\hline
$\overline{6}2m$ &  $\z^{2}$  \\
\hline
$6/mmm$ &  $\z^{3}$  \\
\hline
$23$ &  N/A  \\
\hline
$m\overline{3}$ &  $\z$  \\
\hline
$432$ &  $\zz$  \\
\hline
$\overline{4}3m$ &  $\z$  \\
\hline
$m\overline{3}m$ &  $\z^{2}$  \\
\hline
\end{tabular}
\end{centering}
\caption{Classifications ${\cal C}$ of topological crystalline insulators for non-interacting electrons with time reversal symmetry, significant spin-orbit coupling, and crystalline point group symmetry.  N/A denotes a trivial classification.  Full classifications including strong TIs can be easily obtained from this table by using the group extension rules given in Appendix \ref{sec:TI+TCI}.
\label{tab:PG}}
\end{table}

\begin{table*}
\begin{centering}
\begin{tabular}{ |c|C{1.6cm} |c|C{1.6cm} |c|C{1.6cm} |c|C{1.6cm} |c|C{1.6cm} |c|C{1.6cm}|}
\hline
\#1 & $  \mathbb{Z}_2^3$ &\#41 & $  \mathbb{Z}_2^3$ &\#81 & $  \mathbb{Z}_2^3$ &\#121 & $ \mathbb{Z} \times \mathbb{Z}_2^2$ &\#161 & $  \mathbb{Z}_2$ &\#201 & $  \mathbb{Z}_2^2$ \\
\hline
\#2 & $  \mathbb{Z}_2^4$ &\#42 & $ \mathbb{Z}^2 \times \mathbb{Z}_2^2$ &\#82 & $  \mathbb{Z}_2^2$ &\#122 & $  \mathbb{Z}_2^2$ &\#162 & $ \mathbb{Z} \times \mathbb{Z}_2^2$ &\#202 & $ \mathbb{Z} $ \\
\hline
\#3 & $  \mathbb{Z}_2^4$ &\#43 & $  \mathbb{Z}_2^2$ &\#83 & $ \mathbb{Z}^2 \times \mathbb{Z}_2^2$ &\#123 & $ \mathbb{Z}^5 $ &\#163 & $  \mathbb{Z}_2^2$ &\#203 & $  \mathbb{Z}_2$ \\
\hline
\#4 & $  \mathbb{Z}_2^3$ &\#44 & $ \mathbb{Z}^2 \times \mathbb{Z}_2$ &\#84 & $ \mathbb{Z} \times \mathbb{Z}_2^2$ &\#124 & $ \mathbb{Z} \times \mathbb{Z}_2^3$ &\#164 & $ \mathbb{Z} \times \mathbb{Z}_2^2$ &\#204 & $ \mathbb{Z} \times \mathbb{Z}_2$ \\
\hline
\#5 & $  \mathbb{Z}_2^3$ &\#45 & $  \mathbb{Z}_2^3$ &\#85 & $  \mathbb{Z}_2^3$ &\#125 & $ \mathbb{Z} \times \mathbb{Z}_2^3$ &\#165 & $  \mathbb{Z}_2^2$ &\#205 & $  \mathbb{Z}_2$ \\
\hline
\#6 & $ \mathbb{Z}^2 \times \mathbb{Z}_2^2$ &\#46 & $ \mathbb{Z} \times \mathbb{Z}_2^2$ &\#86 & $  \mathbb{Z}_2^3$ &\#126 & $  \mathbb{Z}_2^3$ &\#166 & $ \mathbb{Z} \times \mathbb{Z}_2^2$ &\#206 & $  \mathbb{Z}_2^2$ \\
\hline
\#7 & $  \mathbb{Z}_2^3$ &\#47 & $ \mathbb{Z}^6 $ &\#87 & $ \mathbb{Z} \times \mathbb{Z}_2^2$ &\#127 & $ \mathbb{Z}^3 \times \mathbb{Z}_2$ &\#167 & $  \mathbb{Z}_2^2$ &\#207 & $  \mathbb{Z}_2^2$ \\
\hline
\#8 & $ \mathbb{Z} \times \mathbb{Z}_2^2$ &\#48 & $  \mathbb{Z}_2^4$ &\#88 & $  \mathbb{Z}_2^2$ &\#128 & $ \mathbb{Z} \times \mathbb{Z}_2^2$ &\#168 & $  \mathbb{Z}_2^2$ &\#208 & $  \mathbb{Z}_2^2$ \\
\hline
\#9 & $  \mathbb{Z}_2^2$ &\#49 & $ \mathbb{Z} \times \mathbb{Z}_2^4$ &\#89 & $  \mathbb{Z}_2^4$ &\#129 & $ \mathbb{Z}^2 \times \mathbb{Z}_2^2$ &\#169 & $  \mathbb{Z}_2$ &\#209 & $  \mathbb{Z}_2$ \\
\hline
\#10 & $ \mathbb{Z}^2 \times \mathbb{Z}_2^3$ &\#50 & $  \mathbb{Z}_2^4$ &\#90 & $  \mathbb{Z}_2^3$ &\#130 & $  \mathbb{Z}_2^3$ &\#170 & $  \mathbb{Z}_2$ &\#210 & $  \mathbb{Z}_2$ \\
\hline
\#11 & $ \mathbb{Z} \times \mathbb{Z}_2^3$ &\#51 & $ \mathbb{Z}^3 \times \mathbb{Z}_2^2$ &\#91 & $  \mathbb{Z}_2^3$ &\#131 & $ \mathbb{Z}^3 \times \mathbb{Z}_2$ &\#171 & $  \mathbb{Z}_2^2$ &\#211 & $  \mathbb{Z}_2^2$ \\
\hline
\#12 & $ \mathbb{Z} \times \mathbb{Z}_2^3$ &\#52 & $  \mathbb{Z}_2^3$ &\#92 & $  \mathbb{Z}_2^2$ &\#132 & $ \mathbb{Z}^2 \times \mathbb{Z}_2^2$ &\#172 & $  \mathbb{Z}_2^2$ &\#212 & $  \mathbb{Z}_2$ \\
\hline
\#13 & $  \mathbb{Z}_2^4$ &\#53 & $ \mathbb{Z} \times \mathbb{Z}_2^3$ &\#93 & $  \mathbb{Z}_2^4$ &\#133 & $  \mathbb{Z}_2^3$ &\#173 & $  \mathbb{Z}_2$ &\#213 & $  \mathbb{Z}_2$ \\
\hline
\#14 & $  \mathbb{Z}_2^3$ &\#54 & $  \mathbb{Z}_2^4$ &\#94 & $  \mathbb{Z}_2^3$ &\#134 & $ \mathbb{Z} \times \mathbb{Z}_2^3$ &\#174 & $ \mathbb{Z}^2 $ &\#214 & $  \mathbb{Z}_2^2$ \\
\hline
\#15 & $  \mathbb{Z}_2^3$ &\#55 & $ \mathbb{Z}^2 \times \mathbb{Z}_2^2$ &\#95 & $  \mathbb{Z}_2^3$ &\#135 & $ \mathbb{Z} \times \mathbb{Z}_2^2$ &\#175 & $ \mathbb{Z}^2 \times \mathbb{Z}_2$ &\#215 & $ \mathbb{Z} \times \mathbb{Z}_2$ \\
\hline
\#16 & $  \mathbb{Z}_2^5$ &\#56 & $  \mathbb{Z}_2^3$ &\#96 & $  \mathbb{Z}_2^2$ &\#136 & $ \mathbb{Z}^2 \times \mathbb{Z}_2$ &\#176 & $ \mathbb{Z} \times \mathbb{Z}_2$ &\#216 & $ \mathbb{Z} $ \\
\hline
\#17 & $  \mathbb{Z}_2^4$ &\#57 & $ \mathbb{Z} \times \mathbb{Z}_2^3$ &\#97 & $  \mathbb{Z}_2^3$ &\#137 & $ \mathbb{Z} \times \mathbb{Z}_2^2$ &\#177 & $  \mathbb{Z}_2^3$ &\#217 & $ \mathbb{Z} \times \mathbb{Z}_2$ \\
\hline
\#18 & $  \mathbb{Z}_2^3$ &\#58 & $ \mathbb{Z} \times \mathbb{Z}_2^2$ &\#98 & $  \mathbb{Z}_2^3$ &\#138 & $ \mathbb{Z} \times \mathbb{Z}_2^2$ &\#178 & $  \mathbb{Z}_2^2$ &\#218 & $  \mathbb{Z}_2$ \\
\hline
\#19 & $  \mathbb{Z}_2^2$ &\#59 & $ \mathbb{Z}^2 \times \mathbb{Z}_2^2$ &\#99 & $ \mathbb{Z}^3 \times \mathbb{Z}_2$ &\#139 & $ \mathbb{Z}^3 \times \mathbb{Z}_2$ &\#179 & $  \mathbb{Z}_2^2$ &\#219 & $  \mathbb{Z}_2$ \\
\hline
\#20 & $  \mathbb{Z}_2^3$ &\#60 & $  \mathbb{Z}_2^3$ &\#100 & $ \mathbb{Z} \times \mathbb{Z}_2^2$ &\#140 & $ \mathbb{Z}^2 \times \mathbb{Z}_2^2$ &\#180 & $  \mathbb{Z}_2^3$ &\#220 & $  \mathbb{Z}_2$ \\
\hline
\#21 & $  \mathbb{Z}_2^4$ &\#61 & $  \mathbb{Z}_2^3$ &\#101 & $ \mathbb{Z} \times \mathbb{Z}_2^2$ &\#141 & $ \mathbb{Z} \times \mathbb{Z}_2^2$ &\#181 & $  \mathbb{Z}_2^3$ &\#221 & $ \mathbb{Z}^3 $ \\
\hline
\#22 & $  \mathbb{Z}_2^4$ &\#62 & $ \mathbb{Z} \times \mathbb{Z}_2^2$ &\#102 & $ \mathbb{Z} \times \mathbb{Z}_2^2$ &\#142 & $  \mathbb{Z}_2^3$ &\#182 & $  \mathbb{Z}_2^2$ &\#222 & $  \mathbb{Z}_2^2$ \\
\hline
\#23 & $  \mathbb{Z}_2^3$ &\#63 & $ \mathbb{Z}^2 \times \mathbb{Z}_2^2$ &\#103 & $  \mathbb{Z}_2^3$ &\#143 & $  \mathbb{Z}_2$ &\#183 & $ \mathbb{Z}^2 \times \mathbb{Z}_2$ &\#223 & $ \mathbb{Z} \times \mathbb{Z}_2$ \\
\hline
\#24 & $  \mathbb{Z}_2^3$ &\#64 & $ \mathbb{Z} \times \mathbb{Z}_2^3$ &\#104 & $  \mathbb{Z}_2^2$ &\#144 & $  \mathbb{Z}_2$ &\#184 & $  \mathbb{Z}_2^2$ &\#224 & $ \mathbb{Z} \times \mathbb{Z}_2^2$ \\
\hline
\#25 & $ \mathbb{Z}^4 \times \mathbb{Z}_2$ &\#65 & $ \mathbb{Z}^4 \times \mathbb{Z}_2$ &\#105 & $ \mathbb{Z}^2 \times \mathbb{Z}_2$ &\#145 & $  \mathbb{Z}_2$ &\#185 & $ \mathbb{Z} \times \mathbb{Z}_2$ &\#225 & $ \mathbb{Z}^2 $ \\
\hline
\#26 & $ \mathbb{Z}^2 \times \mathbb{Z}_2^2$ &\#66 & $ \mathbb{Z} \times \mathbb{Z}_2^3$ &\#106 & $  \mathbb{Z}_2^2$ &\#146 & $  \mathbb{Z}_2$ &\#186 & $ \mathbb{Z} \times \mathbb{Z}_2$ &\#226 & $ \mathbb{Z} \times \mathbb{Z}_2$ \\
\hline
\#27 & $  \mathbb{Z}_2^4$ &\#67 & $ \mathbb{Z}^2 \times \mathbb{Z}_2^3$ &\#107 & $ \mathbb{Z}^2 \times \mathbb{Z}_2$ &\#147 & $  \mathbb{Z}_2^2$ &\#187 & $ \mathbb{Z}^3 $ &\#227 & $ \mathbb{Z} \times \mathbb{Z}_2$ \\
\hline
\#28 & $ \mathbb{Z} \times \mathbb{Z}_2^3$ &\#68 & $  \mathbb{Z}_2^4$ &\#108 & $ \mathbb{Z} \times \mathbb{Z}_2^2$ &\#148 & $  \mathbb{Z}_2^2$ &\#188 & $ \mathbb{Z} \times \mathbb{Z}_2$ &\#228 & $  \mathbb{Z}_2^2$ \\
\hline
\#29 & $  \mathbb{Z}_2^3$ &\#69 & $ \mathbb{Z}^3 \times \mathbb{Z}_2^2$ &\#109 & $ \mathbb{Z} \times \mathbb{Z}_2$ &\#149 & $  \mathbb{Z}_2^2$ &\#189 & $ \mathbb{Z}^3 $ &\#229 & $ \mathbb{Z}^2 \times \mathbb{Z}_2$ \\
\hline
\#30 & $  \mathbb{Z}_2^3$ &\#70 & $  \mathbb{Z}_2^3$ &\#110 & $  \mathbb{Z}_2^2$ &\#150 & $  \mathbb{Z}_2^2$ &\#190 & $ \mathbb{Z} \times \mathbb{Z}_2$ &\#230 & $  \mathbb{Z}_2^2$ \\
\hline
\#31 & $ \mathbb{Z} \times \mathbb{Z}_2^2$ &\#71 & $ \mathbb{Z}^3 \times \mathbb{Z}_2$ &\#111 & $ \mathbb{Z} \times \mathbb{Z}_2^3$ &\#151 & $  \mathbb{Z}_2^2$ &\#191 & $ \mathbb{Z}^4 $ && \\
\hline
\#32 & $  \mathbb{Z}_2^3$ &\#72 & $ \mathbb{Z} \times \mathbb{Z}_2^3$ &\#112 & $  \mathbb{Z}_2^3$ &\#152 & $  \mathbb{Z}_2^2$ &\#192 & $ \mathbb{Z} \times \mathbb{Z}_2^2$ && \\
\hline
\#33 & $  \mathbb{Z}_2^2$ &\#73 & $  \mathbb{Z}_2^4$ &\#113 & $ \mathbb{Z} \times \mathbb{Z}_2^2$ &\#153 & $  \mathbb{Z}_2^2$ &\#193 & $ \mathbb{Z}^2 \times \mathbb{Z}_2$ && \\
\hline
\#34 & $  \mathbb{Z}_2^3$ &\#74 & $ \mathbb{Z}^2 \times \mathbb{Z}_2^2$ &\#114 & $  \mathbb{Z}_2^2$ &\#154 & $  \mathbb{Z}_2^2$ &\#194 & $ \mathbb{Z}^2 \times \mathbb{Z}_2$ && \\
\hline
\#35 & $ \mathbb{Z}^2 \times \mathbb{Z}_2^2$ &\#75 & $  \mathbb{Z}_2^3$ &\#115 & $ \mathbb{Z}^2 \times \mathbb{Z}_2^2$ &\#155 & $  \mathbb{Z}_2^2$ &\#195 & $  \mathbb{Z}_2$ && \\
\hline
\#36 & $ \mathbb{Z} \times \mathbb{Z}_2^2$ &\#76 & $  \mathbb{Z}_2^2$ &\#116 & $  \mathbb{Z}_2^3$ &\#156 & $ \mathbb{Z} \times \mathbb{Z}_2$ &\#196 & N/A && \\
\hline
\#37 & $  \mathbb{Z}_2^3$ &\#77 & $  \mathbb{Z}_2^3$ &\#117 & $  \mathbb{Z}_2^3$ &\#157 & $ \mathbb{Z} \times \mathbb{Z}_2$ &\#197 & $  \mathbb{Z}_2$ && \\
\hline
\#38 & $ \mathbb{Z}^3 \times \mathbb{Z}_2$ &\#78 & $  \mathbb{Z}_2^2$ &\#118 & $  \mathbb{Z}_2^3$ &\#158 & $  \mathbb{Z}_2$ &\#198 & N/A && \\
\hline
\#39 & $ \mathbb{Z} \times \mathbb{Z}_2^3$ &\#79 & $  \mathbb{Z}_2^2$ &\#119 & $ \mathbb{Z} \times \mathbb{Z}_2^2$ &\#159 & $  \mathbb{Z}_2$ &\#199 & $  \mathbb{Z}_2$ && \\
\hline
\#40 & $ \mathbb{Z} \times \mathbb{Z}_2^2$ &\#80 & $  \mathbb{Z}_2^2$ &\#120 & $  \mathbb{Z}_2^3$ &\#160 & $ \mathbb{Z} \times \mathbb{Z}_2$ &\#200 & $ \mathbb{Z}^2 $ && \\
\hline
\end{tabular}
\end{centering}
\caption{Topological crystalline insulator classifications ${\cal C}$ {for non-interacting electrons with time reversal symmetry and significant spin-orbit coupling, given for all space groups. N/A denotes a trivial classification.  Full classifications including strong TIs can be easily obtained from this table by using the group extension rules given in Appendix \ref{sec:TI+TCI}.}}\label{tab:SG}
\end{table*}

\section{Topological crystals beyond layer construction}
\label{app:beyond}

To find the topological crystals that cannot be decomposed into decoupled planar layers, we can focus on $\zz$ TCIs, because all MTCIs have a layer construction.  
We then find these states by identifying those space groups where our classification of $\zz$ TCIs is larger than the number of 2dTI layer constructions with distinct invariants tabulated in \cite{Song2017a}.  
This occurs only for twelve space groups, and when it occurs, the discrepancy is always a factor of two, meaning there is a single $\zz$ factor generated by a non-layer-constructible state.  
It is then straightforward to calculate $\delta(g)$ for these states, because $\delta(g)$ is given in terms of invariants for the 2dTI layer  states \cite{Song2017a}. 
So we just need to find $\delta(g)$ that cannot be obtained by stacking the layer states.
In Fig.~\ref{fig:nonLC} we plotted all these twelve topological crystals, and in Table~\ref{tab:beyond} we list the $\zz$ invariants of these states.
Here we give the coordinates of the inequivalent 2-cells plotted in Fig.~\ref{fig:nonLC}. 
\begin{enumerate}
\item \#34. The corners of the first decorated 2-cell are $(1,\frac12,\frac34)$, $(1,0,\frac34)$, $(\frac12,0,1\frac14)$, $(\frac12,\frac12,1\frac14)$, and the corners of the second decorated 2-cell are $(1,0,1\frac34)$,  $(\frac12,0,1\frac14)$, $(1,0,\frac34)$, $(1\frac12,0,1\frac14)$.
\item \#48. The first 2-cell $(\frac14,\frac34,0)$, $(\frac14,\frac14,0)$, $(\frac34,\frac14,0)$, $(\frac34,\frac34,0)$, the second 2-cell $(\frac14,\frac34,0)$, $(\frac14,\frac14,0)$,  $(\frac14,\frac14,\frac12)$, $(\frac14,\frac34,\frac12)$, and the third 2-cell $(\frac34,\frac34,\frac12)$, $(\frac14,\frac34,\frac12)$, $(\frac14,\frac34,0)$, and $(\frac34,\frac34,0)$.
\item \#77. The first 2-cell $(0,\frac12,\frac14)$, $(0,0,0)$, $(0,0,\frac12)$, the second 2-cell $(\frac12,0,\frac34)$, $(\frac12,\frac12,1)$, $(1,\frac12,1\frac14)$, $(1,0,1)$, and the third 2-cell $(0,\frac12,\frac14)$, $(\frac12,\frac12,\frac12)$, $(1,\frac12,\frac14)$, $(\frac12,\frac12,0)$.
\item \#86. The first 2-cell $(1\frac14,\frac14,\frac34)$, $(\frac34,\frac14,1)$, $(\frac34,\frac34,1\frac14)$, and the secnd 2-cell $(\frac34,\frac34,\frac14)$, $(\frac14,1\frac14,0)$, $(\frac34,1\frac34,\frac14)$, $(\frac34,1\frac14,\frac12)$.
\item \#93. The first 2-cell $(1,0,\frac12)$, $(\frac12,\frac12,\frac12)$, $(\frac12,\frac12,0)$, $(1,0,0)$, and the second 2-cell $(\frac12,\frac12,0)$, $(1,0,0)$, $(1\frac12,\frac12,0)$.
\item \#94. The first 2-cell $(0,0,0)$, $(0,0,\frac12)$, $(0,\frac12,\frac12)$, $(0,\frac12,0)$, and the second 2-cell $(0,\frac12,\frac12)$, $(0,0,\frac12)$, $(\frac12,0,\frac12)$, $(\frac12,\frac12,\frac12)$.
\item \#102. The first 2-cell $(1,0,\frac34)$, $(\frac12,\frac12,\frac14)$, $(1,\frac12,\frac12)$, and the second 2-cell $(\frac12,\frac12,\frac14)$, $(1,\frac12,0)$, $(1\frac12,\frac12,\frac14)$, $(1,\frac12,\frac12)$.
\item \#118. The first 2-cell $(0,\frac12,\frac34)$, $(0,\frac12,\frac14)$, $(0,0,\frac14)$, $(0,0,\frac34)$, and the second 2-cell $(0,\frac12,\frac14)$, $(0,0,\frac14)$, $(\frac12,0,\frac14)$, $(\frac12,\frac12,\frac14)$.
\item \#134. The first 2-cell $(\frac12,0,\frac14)$, $(\frac12,0,\frac34)$, $(\frac14,\frac14,\frac34)$, $(\frac14,\frac14,\frac14)$, and the second 2-cell $(\frac12,0,\frac14)$, $(\frac34,\frac14,\frac14)$, $(\frac12,\frac12,\frac14)$, $(\frac14,\frac14,\frac14)$.
\item \#201. The first 2-cell $(\frac14,\frac14,\frac34)$, $(\frac12,\frac12,\frac12)$, $(\frac34,\frac14,\frac34)$.
\item \#208. The first 2-cell $(0,\frac12,0)$, $(\frac14,\frac34,\frac14)$, $(\frac12,\frac12,0)$, and the second 2-cell $(0,\frac12,0)$, $(0,\frac12,\frac12)$, $(\frac14,\frac34,\frac14)$.
\item \#224. The first 2-cell $(\frac12,\frac12,1)$, $(0,\frac12,\frac12)$, $(\frac14,\frac34,\frac34)$.
\end{enumerate}
Equivalent 2-cells can be obtained by acting the symmetries listed in Table.~\ref{tab:beyond} on the above 2-cells.

In Table~\ref{tab:beyond}, we also list the symmetry-based indicators, when they exist, of the non-layer-constructible TCIs.
Only five of the twelve space groups have nontrivial indicator groups, and, all the corresponding indicators ($z_{2w,1},z_{2w,2},z_{2w,3},z_{4}$) are determined entirely by the inversion eigenvalues \cite{Song2017a,Khalaf17symmetry}.
In other words, the indicators of these TCIs remain unchanged as we break the symmetry down to space group $P\bar{1}$ (\#2), where the only remaining point group symmetry is inversion.
On the other hand, for TCIs in space group $P\bar{1}$, it is known that the $z_{2w,i=1,2,3}$ indicators are equivalent to the $\mathbb{Z}_2$ weak invariants, and the $z_4$ indicator is equivalent to the $\mathbb{Z}_2$ inversion invariant ($z_4 = 2 \delta(\{\overline{1}|000\})$) \cite{Song2017a}.
Therefore, the indicators in these five space groups can be directly read out from the topological invariants.

\begin{table*}
\begin{centering}
\begin{tabular}{|c|C{6.7cm}|C{8.3cm}|c|}
\hline
SG & \multicolumn{2}{|c|}{Invariants} & Indicator \\
\hline
\#34 & $\delta(\{1|100\})=1$, $\delta(\{1|010\})=1$, $\delta(\{1|001\})=1$ & $\delta(\{2_{001}|000\})=0$, $\delta(\{m_{100}|\frac12\frac12\frac12\})=0$ & N/A \\
\hline
\#48 & $\delta(\{1|100\})=1$, $\delta(\{1|010\})=1$, $\delta(\{1|001\})=1$ & $\delta(\{2_{001}|\frac12\frac12 0\})=0$, $\delta(\{2_{100}|0\frac12\frac12\})=0$, $\delta(\{\overline{1}|000\})=1$ & 1112 \\
\hline
\#77 & $\delta(\{1|100\})=1$, $\delta(\{1|010\})=1$, $\delta(\{1|001\})=1$ & $\delta(\{4_{001}|00\frac12\})=0$ & N/A \\
\hline
\#86 & $\delta(\{1|100\})=1$, $\delta(\{1|010\})=1$, $\delta(\{1|001\})=1$ & $\delta(\{4_{001}|0\frac12\frac12\})=1$, $\delta(\{\overline{1}|000\})=1$ & 1112\\
\hline
\#93 & $\delta(\{1|100\})=1$, $\delta(\{1|010\})=1$, $\delta(\{1|001\})=1$ & $\delta(\{4_{001}|00\frac12\})=0$, $\delta(\{2_{100}|000\})=0$ & N/A\\
\hline
\#94 & $\delta(\{1|100\})=0$, $\delta(\{1|010\})=0$, $\delta(\{1|001\})=1$ & $\delta(\{4_{001}|\frac12\frac12\frac12\})=0$, $\delta(\{2_{110}|000\})=0$ & N/A\\
\hline
\#102 & $\delta(\{1|100\})=1$, $\delta(\{1|010\})=1$, $\delta(\{1|001\})=1$ & $\delta(\{4_{001}|\frac12\frac12\frac12\})=1$, $\delta(\{m_{100}|\frac12\frac12\frac12\})=1$ & N/A\\
\hline
\#118 & $\delta(\{1|100\})=1$, $\delta(\{1|010\})=1$, $\delta(\{1|001\})=1$ & $\delta(\{4_{001}|000\})=0$, $\delta(\{m_{100}|\frac12\frac12\frac12\})=1$ & N/A\\
\hline
\#134 & $\delta(\{1|100\})=1$, $\delta(\{1|010\})=1$, $\delta(\{1|001\})=1$ & $\delta(\{4_{001}|\frac12 0\frac12\})=1$, $\delta(\{2_{100}|0\frac12\frac12\})=0$, $\delta(\{\overline{1}|000\})=0$ & 1110\\
\hline
\#201 & $\delta(\{1|100\})=1$, $\delta(\{1|010\})=1$, $\delta(\{1|001\})=1$ & $\delta(\{2_{001}|\frac12\frac12 0\})=0$, $\delta(\{3_{111}|000\})=0$, $\delta(\{\overline{1}|000\})=1$ & 1112\\
\hline
\#208 & $\delta(\{1|100\})=1$, $\delta(\{1|010\})=1$, $\delta(\{1|001\})=1$ & $\delta(\{4_{001}|\frac12\frac12\frac12\})=1$, $\delta(\{3_{111}|000\})=0$ & N/A\\
\hline
\#224 & $\delta(\{1|100\})=1$, $\delta(\{1|010\})=1$, $\delta(\{1|001\})=1$ & $\delta(\{4_{001}|0 \frac12\frac12\})=1$, $\delta(\{3_{111}|000\})=0$, $\delta(\{\overline{1}|000\})=1$ & 1112\\
\hline
\end{tabular}
\end{centering}
\caption{Topological crystals beyond layer construction. In the second and third columns $\delta(g)$ is given by its values on the generators of the space group. In the last column, symmetry-based indicators ($z_{w,1},z_{w,2},z_{w,3},z_4$) \cite{Song2017a} of centrosymmetric space groups are tabulated.}\label{tab:beyond}
\end{table*}

\section{Unified classification of TI and TCI with space group and time-reversal symmetries} \label{sec:TI+TCI}
\label{app:unified}

We have focused thus far on classifying TCIs, where crystalline symmetry is required to protect a non-trivial cSPT phase.  Here, we address a more general problem, namely the classification of all $d=3$ free-electron insulators with time reversal symmetry, significant spin-orbit coupling, and arbitrary crystalline point group or space group symmetry.  We still ignore distinctions among atomic insulators, so the one new state that must be added as a generator of the classification is the strong topological insulator (STI), which is of course robust even upon breaking crystalline symmetry.

First, we assume that the STI is compatible with an arbitrary crystalline symmetry.  We expect that this is true, but to our knowledge it has not been proved rigorously.  One argument in favor of this expectation is to note that the STI can be described by a continuum theory of a massive Dirac fermion, which is invariant under arbitrary rigid motions of three-dimensional space.  This symmetry can be broken down to an arbitrary space group or point group symmetry, for instance by adding a periodic potential, which produces a model of an STI with arbitrary space group symmetry.  This is not quite a rigorous argument because one has to show that it is possible to regularize the continuum theory in a manner compatible with an arbitrary lattice symmetry.  Another argument is to note that any centrosymmetric space group has a $\mathbb{Z}_4$ indicator\cite{Po2017} and according to the Fu-Kane formula\cite{Fu2007}, the root state with $z_4=1$ is an STI. While it was argued that any symmetry indicator can be realized by a band structure, there is no guarantee the resulting band structure is an insulator \cite{Po2017}.  Assuming an STI can indeed be found for each centrosymmetric space group, then we need only note that every non-centrosymmetric space group is a subgroup of some centrosymmetric space group, so an STI compatible with the latter is compatible with the former.  To show this expectation holds rigorously, a straightforward approach would be to find a small number of space groups that contain all space groups as subgroups, and exhibit a model realizing an STI for each of these symmetry groups.

Next, we would like to compute the classification ${\cal C}_{{\rm full}}$ including both TCIs and STIs.  The topological crystal picture tells us that TCIs are a subgroup (\emph{i.e.} ${\cal C} \subset {\cal C}_{{\rm full}}$), because stacking two TCIs produces another TCI or a trivial state.  It is also true that ${\cal C}_{{\rm full}} / {\cal C} \simeq \zz$, because this quotient corresponds to ignoring the distinctions among TCIs, which leaves only a $\zz$ generated by the STI.  It follows that $| C_{{\rm full}} | = 2 | {\cal C}|$ when the TCI classification is finite.   Na\"{\i}vely one might expect ${\cal C}_{{\rm full}} = {\cal C} \times \zz$, with the $\zz$ factor generated by the STI, but this is not true in general because stacking two identical STIs can result in a non-trivial TCI.  Put another way, ${\cal C}_{{\rm full}}$ can be a non-trivial group extension of ${\cal C}$ by $\zz$, and we need to solve this group extension problem.

We proceed by choosing a particular strong topological insulator state, and stacking this state with itself to get a state we call $(STI)^2$.  We know that $(STI)^2$ has trivial strong index and is thus either a non-trivial TCI or a trivial state; that is, $(STI)^2 \in {\cal C}$.  We need to determine the element of ${\cal C}$ given by $(STI)^2$.  First, we observe that our choice of $STI$ is arbitrary under stacking with a TCI, because such stacking does not change the strong invariant.  It is obvious that stacking $STI$ with a $\zz$ TCI does not affect $(STI)^2$.  But stacking $STI$ with a MTCI can change the $\z$ invariants of $(STI)^2$ by arbitrary even integers, depending on the choice of MTCI.  We thus see that the information in $(STI)^2$ that is independent of the arbitrary choice of $STI$ is \emph{precisely} the information preserved under the map $\pi : {\cal C} \to \tC \simeq H^1(G_c, \zz)$ introduced in Appendix~\ref{app:H1}.  Therefore, $(STI)^2$ is characterized by a homomorphism from $G_c \to \zz$, namely $\pi( (STI)^2 ) \in H^1(G_c, \zz)$.  Determining $\pi( (STI)^2 )$ solves the group extension problem and determines the group ${\cal C}_{{\rm full}}$.

Denoting the homomorphism given by $\pi( (STI)^2 )$ by $\delta : G_c \to \zz$, it is natural to conjecture that $\delta(g) = 0$ when $g$ is a rigid motion preserving the orientation of space (\emph{e.g.} translations and rotations), and $\delta(g) = 1$ when $g$ reverses orientation (\emph{e.g.} inversion, reflections, glide reflections).  This conjecture is natural because the map $\delta$ should depend only on the crystalline symmetry $G_c$, and there does not seem to be any other non-trivial map that can be defined in a uniform way for all $G_c$.

We can establish this conjecture using results of Khalaf \emph{et. al.} \cite{Khalaf17symmetry}, where the authors studied surface theories obtained by stacking two STIs.  For a crystalline symmetry $G^{{\rm surf}}_c$ preserved by some surface termination, they considered a mass texture on the boundary satisfying $m_{g {\bf r}} = s_g m_{{\bf r}}$, where $g \in G^{{\rm surf}}_c$, ${\bf r}$ is a point on the boundary, $m_{{\bf r}}$ is the Dirac mass, and $s_g = \pm 1$ keeps track of sign changes in the mass.  They showed that $s_{g_1 g_2} = s_{g_1} s_{g_2}$, \emph{i.e.} $s_g$ defines a homomorphism from $G^{{\rm surf}}_c$ to $\zz$.  Moreover, they showed that, in the case of stacking two \emph{identical} STIs,  $s_g = \det R_g$, where $R_g$ is   (This result follows from Eq.~14 of \cite{Khalaf17symmetry} upon taking $\eta^{(1)}_g = \eta^{(2)}_g$, as appropriate for identical STIs.)  Here $g$ is the rigid motion $\{ R_g | \mathbf{t}_g \}$, where $R_g$ is an ${\rm O}(3)$ matrix and $\mathbf{t_g}$ is a translation vector.  Because $\det R_g =  1$ ($\det R_g = -1$) for orientation-preserving  (orientation-reversing) operations, this result is identical to our conjecture upon identifying $s_g = (-1)^{\delta(g)}$.  Physically, $s_g$ and $\delta(g)$ should be thus identified, because the gapless lines on the surface where the mass changes sign are, in the topological crystal picture, precisely the gapless edges of 2-cells touching the surface.

The argument is not quite complete, because the crystalline symmetry $G_c$ cannot generally be preserved by a surface termination.  However, we found that specifying the seven types of invariants listed in the main text uniquely determines a TCI phase (element of ${\cal C}$).  Therefore, we can take $G^{{\rm surf}}_c$ to be the subgroup of $G_c$ associated with each invariant.  It is always possible to choose a surface termination preserving such $G^{{\rm surf}}_c$, so we can run the above argument for each such subgroup. This then determines $\pi( (STI)^2 )$.

This result determines the group structure of ${\cal C}_{{\rm full}}$.  There are three cases:  (i)  If $G_c$ contains only orientation-preserving operations, then ${\cal C}_{{\rm full}} = {\cal C} \times \zz$.  (ii) If $G_c$ contains orientation-reversing operations but no mirror reflections, and hence ${\cal C}$ has no $\z$ factors, then one of the $\zz$ factors in ${\cal C}$ is replaced in ${\cal C}_{{\rm full}}$ by a $\z_4$ factor.  (iii) If $G_c$ contains mirror reflections, then $(STI)^2$ generates a $\z$ factor in ${\cal C}$.  In this case ${\cal C}_{{\rm full}}$ and ${\cal C}$ have the same group structure, but in ${\cal C}_{{\rm full}}$ the generator of one of the $\z$ factors is a STI state.  These rules easily allow one to obtain ${\cal C}_{{\rm full}}$ for all the crystallographic point groups and space groups.  Because $\pi ( (STI)^2 )$ is known, it is also straightforward to explicitly construct the topological crystal corresponding to $(STI)^2$, up to the arbitrariness in defining $STI$.

\end{document}